\documentclass[conf]{new-aiaa}
\usepackage[utf8]{inputenc}

\usepackage{color}
\usepackage{graphicx}
\usepackage{amsmath}
\usepackage[version=4]{mhchem}
\usepackage{siunitx}
\usepackage{longtable,tabularx}
\usepackage{caption}
\usepackage{float}
\usepackage[nameinlink,capitalise,noabbrev]{cleveref}

\usepackage{cancel}
\usepackage{algorithm}
\usepackage{algorithmic}
\graphicspath{{Figures/}}
\usepackage{subfigure}
\usepackage{multirow}

\usepackage{xcolor}
\usepackage{booktabs}

\usepackage{arydshln}

\graphicspath{{Figures/}, {Figures/Syntetic_aero}, {Figures/2D_test_case},  {Figures/3D_test_case},  {Figures/3D_test_case/harmonic_signals}, {Figures/3D_test_case/pitch_step_response}}

\title{Parametric Nonlinear Volterra Series via Machine Learning: Transonic Aerodynamics}

\author{Gabriele Immordino\footnote{PhD Student.}, Andrea Da Ronch\footnote{Professor, AIAA Senior Member.}
}
\affil{Faculty of Engineering and Physical Sciences\\University of Southampton, Southampton, United Kingdom, SO17~1BJ}
\author{Marcello Righi\footnote{Professor, AIAA Member, Lecturer at Federal Institute of Technology Zurich ETHZ}}
\affil{School of Engineering\\Zurich University of Applied Sciences ZHAW, Winterthur, Switzerland, 8401 
}

\begin{document}

\maketitle

\begin{abstract}

This study introduces an approach for modeling unsteady transonic aerodynamics within a parametric space, using Volterra series to capture aerodynamic responses and machine learning to enable interpolation. The first- and second-order Volterra kernels are derived from indicial aerodynamic responses obtained through computational fluid dynamics, with the second-order kernel calculated as a correction to the dominant linear response. Machine learning algorithms, specifically artificial neural network and Gaussian process regression, are used to interpolate kernel coefficients within a parameter space defined by Mach number and angle of attack. The methodology is applied to two and three dimensional test cases in the transonic regime. Results underscore the benefit of including the second-order kernel to address strong nonlinearity and demonstrate the effectiveness of neural networks. The approach achieves a level of accuracy that appears sufficient for use in conceptual design.

\end{abstract}

\section*{Nomenclature}
\textbf{Acronyms} 
{\renewcommand\arraystretch{1.0}
\noindent\begin{longtable*}{@{}l @{\quad=\quad} l@{}}

$BSCW$ & Benchmark Super Critical Wing \\
$CFD$ & Computational Fluid Dynamics \\
$FCNN$ & fully--connected neural network \\
$GPR$ &  Gaussian process regression \\
$LHS$ & Latin Hypercube Sampling \\
$ML$ & machine learning \\
$MSE$ & mean squared error \\
$ROM$ & reduced--order model \\
$SVD$ & Singular Value Decomposition  \\

\end{longtable*}}

\textbf{Symbols} 
{\renewcommand\arraystretch{1.0}
\noindent\begin{longtable*}{@{}l @{\quad=\quad} l@{}}

$A^+$ & pseudo-inverse of the matrix $A$ \\
$\alpha_0$   & freestream angle of attack, deg \\
$\alpha$   & pitch angle, deg \\
$\bar{\alpha}$   & mean pitch angle, deg \\
$\alpha_A$    &  angular amplitude, deg \\
$c$   & chord \\
$C_{L}$ & lift coefficient \\
$C_{M}$ & pitching moment coefficient \\
$C_P/deg$ & pressure coefficient normalized by oscillation amplitude \\
$(C_P)_{Im}/deg$ & imaginary component of pressure coefficient normalized by oscillation amplitude \\
$(C_P)_{mean}$ & mean pressure coefficient \\
$(C_P)_{Re}/deg$ & real component of pressure coefficient normalized by oscillation amplitude \\
$f$ & frequency, Hz \\
$h()$ & convolution kernels \\
$h_0 ... h_n$ & Volterra kernels \\
$H_n$ & Volterra operator of order n \\
$L_r()$ & Laguerre polynomials of order r \\
$m_1 ... m_M$ & memory lag terms \\
$M$ & Mach number \\
$Re$ & Reynolds number \\
$\theta_r$ & weights of the Laguerre polynomials \\
$\tau$ & reduced time \\

\end{longtable*}}

\section{Introduction} \label{sec_introduction}

In aerospace and mechanical engineering, the design process for new products relies on hierarchies of mathematical models, the physical complexity of which may be imposed by computational costs or dictated by regulations. These models typically incorporate parameters to account for various operating conditions and configurations. In the framework of optimization, for example, hundreds of parameters (design variables) may be required to define the configuration of a system. Similarly, uncertainty propagation may necessitate defining a complex parameter space to account for variations in geometrical imperfections, material properties, or flow conditions.

The design process, especially  during optimization and uncertainty quantification, often involves numerous evaluations of the system's mathematical models across a wide range of points in the parameter space. Computational costs vary with the level of model fidelity: lower fidelity models are traditionally used for computationally intensive evaluations, while higher fidelity models -- often involving nonlinear partial differential equations discretized on fine grids -- are typically reserved for later stages of the design process. Additionally, the type of analysis significantly impacts computational demands. For example, evaluating a system's aeroelastic response requires either time-accurate simulations or high-resolution discretization in the frequency domain, both of which can be computationally expensive.

Reduced-order models (ROMs) offer the possibility to retain, in principle, the same accuracy provided by a full model, at much lower computational costs. Many review papers are available for each engineering field. The reader is referred to Refs.~\cite{dowell2023reduced,silva1991methodology} for a more multifaceted and nuanced discussion of the topic. The capability of ROMs to capture nonlinear behavior is often a key factor in selecting the appropriate model for a given application.
Traditional ROMs often utilize techniques such as proper orthogonal decomposition (POD)~\cite{chatterjee2000introduction} or dynamic mode decomposition (DMD)~\cite{schmid2010dynamic} to project higher-dimensional manifolds onto lower-dimensional spaces. Although effective for linear phenomena~\cite{lassila2014model, lucia2004reduced}, they may be insufficient to reconstruct nonlinear effects, typically those appearing in the aerodynamic response of wings and aerofoils in the transonic regime, which is often considered the ultimate benchmark.  

ROMs can be parametric, with their development often leveraging coefficient interpolation, at least within certain parameter ranges space~\cite{amsallem2008interpolation,amsallem2009method,amsallem2012nonlinear,zahr2015progressive}. Ensuring a certain smoothness in the parameter space is often regarded as a key prerequisite for efficient interpolation of coefficients~\cite{amsallem2011online}. Over the past few decades, a substantial body of literature has emerged on this topic~\cite{shaw1999modal,hall2002computation,schuster2003computational,lucia2004reduced,lassila2014model,dowell2023reduced}. The advent of machine learning (ML) techniques offers valuable support due to their ability to detect patterns. The notion that a neural network projects information onto a latent space where it can be manipulated efficiently may be interpreted as a particular implementation of the recommendations in Ref.~\cite{amsallem2011online} regarding the projection of ROM coefficients onto a smooth manifold.

Volterra series are especially used to model unsteady responses. Among the many ROM approaches proposed over the last decades, Volterra series holds significant prominence in several engineering fields~\cite{silva1999reduced}. Volterra series~\cite{volterra1887sopra} are the multi--dimensional extension of the impulse response concept from linear system theory, allowing the use of Volterra kernels for the characterization of nonlinear systems. The method consists of a series of kernels that capture the various nonlinear interactions between the system's inputs and outputs. Each kernel in the series corresponds to a specific degree of nonlinearity, progressively representing higher-order effects. One of the main challenges in applying the Volterra series lies in the precise identification of the kernel coefficients in particular for higher order kernels, which can require a large amount of data and significant computational resources~\cite{dodd2002new,koh1985second}.

Several techniques have been proposed to address these problems~\cite{silva1999reduced}. Marzocca et al.\cite{marzocca2004nonlinear} investigated the use of Volterra series from aerodynamic indicial functions to study the aeroelastic response and flutter prediction of simple aeroelastic systems in incompressible flow, proposing a promising approach for addressing nonlinear aeroelasticity challenges. Loghmanian et al.\cite{loghmanian2011nonlinear} employed a genetic algorithm to optimize the structure of Volterra kernels for modeling nonlinear discrete-time systems. Da Silva et al.\cite{shiki2023practical} advanced this work by integrating a discrete-time Volterra model with orthonormal Kautz functions, enabling the identification of nonlinearities in the system through the analysis of contributions from both linear and nonlinear Volterra kernels. Nie and Wu\cite{lianrui2023matched} introduced the Matched Volterra ROM, improving prediction accuracy for transonic aerodynamic loads by matching step motions in equal intervals, particularly for large-amplitude oscillations. Israelsen and Smith~\cite{israelsen2014generalized} reduced the number of unknowns and simplifies the identification process by projecting the Volterra kernels onto a subspace defined by Laguerre polynomials. ML techniques have also been employed for Volterra kernel identification~\cite{de2019multi,de2019volterra,ross2021learning,wray1994calculation}, offering an innovative pathway for enhancing the efficiency of kernel determination. Recently, Levin et al.~\cite{levin2022convolution} proposed a two-step identification process: whereas the linear kernel is identified from a small amplitude input signal, the higher order kernels are a mere "correction", representing the deviations from the linear behaviour, and are identified from larger amplitude inputs. The concept builds on earlier research by Balajewicz and Dowell~\cite{balajewicz2012reduced}, which explored the application of the Sparse Volterra Series for reduced-order modeling of flutter and limit-cycle oscillations in dynamic systems. Their findings demonstrated that this approach effectively captured the system's nonlinearities and memory effects, while also achieving a significant reduction in computational costs.

In this study, we focus on the identification of Volterra kernels at selected points within a parameter space and on the subsequent reconstruction or interpolation of the kernel coefficients across a larger set of points. While the parameter space in our examples is defined by Mach number ($M$) and freestream angle of attack (\(\alpha_0\)), the methodology is versatile and can be applied to deliver the unsteady aerodynamic response for any combination of parameters, making it well-suited for optimization or flutter tracking along specific parameter paths, such as constant $M$ or constant \(\alpha_0\). To determine the Volterra kernel coefficients, we adopted the approach of Levin et al.~\cite{levin2022convolution}, which separates the identification of linear and nonlinear components of the system's response. This method proves pragmatic and robust, as the linear kernels, which typically dominate the system, are uniquely determined. For interpolation, we explored two ML techniques, Gaussian Process Regression and Artificial Neural Network, with the latter demonstrating superior accuracy in nearly all experiments. The effectiveness of the Volterra ROM is demonstrated through three test cases: a synthetic aerodynamic model, a 2D aerofoil, and a 3D wing, focusing on transonic conditions.

The paper continues in Section~\ref{sec_methodology} with a detailed description of the proposed methodology. The following three sections—Sections~\ref{sec:synthetic_aero} to \ref{sec_bscw}—present results from test cases of increasing complexity, ranging from synthetic aerodynamic data to three-dimensional results for a complex wing configuration, highlighting the method's versatility across diverse applications. Concluding remarks are provided in Section~\ref{sec_conclusions}.

\section{Methodology} \label{sec_methodology}

The methodology is designed around Volterra series for a single-input single-output (SISO) system. 
First, we outline the approach for identifying the linear kernel independently from the higher-order kernels. Next, we introduce two ML techniques to construct a parametric Volterra model for a SISO system, with dependencies on \(\alpha_0\) and $M$. Although this study focuses on SISO systems, the proposed approach and methods are extendable to multi-input systems. It is worth noting that kernel identification in such cases is discussed less extensively in the literature, as pointed out by Worden~\cite{worden1997harmonic}.

\subsection{Volterra series for a SISO system}

Given a discrete-time, time-invariant and nonlinear system with single-input $u$, the output $y(n)$ can be approximated be means of the $p$-th order Volterra series: 

\begin{equation}
    y (n) = \sum_{i=1}^p H_i [u(n)],
    \label{general_volterra_series}
\end{equation}

where the Volterra series is composed by $p$ operators, $H_i$, each with a memory depth $m$ that is independent from $p$.  The first  operator is as follows: 

\begin{equation}
    H_1[u(n)] = \sum_{k=1}^m h_1(k) \, u(n-k), 
\label{eq:1st_order_volterra}
\end{equation}
and corresponds to a linear convolution operator, which can be identified through indicial responses\cite{righi2019subsonic}. The following two operators are of the form: 
\begin{equation}
    H_2[u(n)] = \sum_{k_1=1}^m \sum_{k_2=1}^m h_2(k_1,k_2) \, u(n-k_1)\,u(n-k_2), 
\label{eq:2nd_order_volterra}
\end{equation}

\begin{equation}
    H_3[u(n)] = \sum_{k_1=1}^m \sum_{k_2=1}^m \sum_{k_3=1}^m h_3(k_1,k_2,k_3) \, u(n-k_1)\,u(n-k_2)\,u(n-k_3), 
\end{equation}
In general, the $i$-th operator, $H_i$, represents the $i$-th order discrete-time convolution between the input $u$ and the Volterra kernels $h_i$: 
\begin{equation}
    H_i [u(n)] = \sum_{k_1 = 1}^m \dots \sum_{k_i = 1}^m h_i(k_1, \dots, k_i, \dots) \prod_{z=1}^i u(n - k_z), 
\end{equation}
where, in most cases, the kernels are symmetrical with respect to any permutation of the indices, i.e. $h(k_1, k_2) = h(k_2, k_1)$. For a system with multiple inputs, $u_i$, the Volterra operators are extended to take into account all possible combinations of the inputs, see for example~\cite{balajewicz2012reduced,balajewicz2010application}. 

It is well known that increasing the order of the Volterra series introduces a large number of kernels, the identification of which may become demanding or prohibitive if no simplifications are introduced as pointed out by Levin et al. in Ref.~\cite{levin2022convolution}. In practice, a well-known assumption consists in removing cross-memory effects, $h(k_i,\,k_j) = 0$ for $i \ne j$, as presented by Balajewicz and Dowell~\cite{balajewicz2012reduced} for applications similar to those presented herein. It is worth noting that data regularization is also exploited to reduced the number of coefficients as mentioned by Cheng et al.~\cite{cheng2017volterra}. This means, in practice, that kernels of a given operator are expanded in terms of a set of basis functions, typically Laguerre polynomials (refer to~\cite{campello2004optimal} for a thorough presentation). For instance, the first order kernel for a SISO system is represented by a series expansion with the first $R$ Laguerre polynomials: 

\begin{equation} \label{eq:kernels_laguerre}
h_1(j)   = \sum_{r=1}^R \theta_r \, L_r (t_j),
\end{equation}
which means that the $j$-th kernel is obtained from the combination of $R$ Laguerre polynomials evaluated at the time $t_j$. The first order output can therefore be expressed as: 

\begin{equation}
    y (t) = \sum_{k=1}^m \sum_{r=1}^R \theta_r \, L_r (t_k) \, u(n-k),
    \label{eq:volterra_laguerre1}
\end{equation}

The order of the Laguerre polynomials $R$ is set on a case by case basis. 
In the present study, a number of kernels up to several hundreds was used, whereas only ten to twenty Laguerre polynomials were sufficient to obtain a level of accuracy deemed acceptable. Moreover, the regularization provided by the Laguerre polynomials also provides additional robustness in the identification process. In the context of Laguerre polynomials, it is necessary to point out that a series expansion is in principle incompatible with a parametric realization of the Volterra series, which means that the kernels are reconstructed over a parameter space (the expression ``interpolation'' may be used to be consistent with the literature, even though different techniques are exploited in this study). As mentioned by Amsallem and Farhat~\cite{amsallem2008interpolation}, a set of interpolated Laguerre coefficients $\theta_r$, may not provide a consistent expansion of the interpolated kernels. As such, this study exploits Laguerre polynomials primarily for data regularization but not for reconstructing the kernels over a parametrs space. 

\subsubsection{Identification of Linear Volterra kernel for a SISO System}

Herein, a step function or a smoothed step function were used for all test cases. When using the linear Volterra kernels only, the approximated response is equivalent to a conventional convolutional integral in discrete time: 

\begin{equation}
    y(n) = \sum_{k=1}^m h_1(k) \, u(n-k),
\label{eq:linear_response_scalar}    
\end{equation}

Eq.~\ref{eq:linear_response_scalar} can be written in matrix form for $n$ values of the response:  

\begin{equation}
    \{ y \} = [U] \{h\},  
    \label{eq:convolution_1}
\end{equation}

where $U$ is $n \times m$ matrix (a lower triangular square matrix if $n = m$) containing the input values re-arranged so that the input signal starting from the row corresponding to the number of the column:

\begin{equation}
    U(i,j) = u(i - j),
\end{equation}

also expressed in matrix form: 

\begin{equation}
    [U] = \left[
    \begin{array}{ccccc}
       u(0)   &        &        &        &  \\
       u(1)   & u(0)   &        &        &  \\ 
       u(2)   & u(1)   & u(0)   &        &  \\ 
       \vdots & \vdots & \vdots &        &  \\ 
       u(m)   & u(m-1) & u(m-2) & \dots  & u(0)  \\ 
       \vdots & \vdots & \vdots & \ddots & \vdots \\ 
       u(n)   & u(n-1) & u(n-2) & \dots  & u(n-m)  \\ 
    \end{array}
    \right].
\label{eq:inputugenral}    
\end{equation}

Note that in case a step input signal of magnitude $\alpha$ is used, $u(i \ge 0) = \alpha$ and $U$ has the form:

\begin{equation}
    [U] = \alpha \, \left[
    \begin{array}{ccccc}
       1  &  0 & 0 & \dots & 0 \\
       1  &  1 & 0 & \dots & 0 \\ 
       \vdots & \vdots & \vdots & \ddots & \vdots \\ 
       1 & 1 & 1 & 1 & 1 \\
 
       \vdots & \vdots & \vdots & \vdots & \vdots \\ 
       1 & 1 & 1 & 1 & 1 \\ 
    \end{array}
    \right].
\label{eq:inputustep}    
\end{equation}

The linear kernels are then obtained by inverting Eq.~\ref{eq:convolution_1}:

\begin{equation}
    \{ h \} = [U]^{+} \{y\}.
    \label{eq:linearsolution_1}
\end{equation}

where $[U]^{+}$ denotes the pseudo-inverse of $[U]$ if $n>m$ and the inverse of $[U]$ if $n=m$. The pseudo-inverse of a matrix $[A]$ (shape $n\times m$) is obtained from Singular Value Decomposition (SVD): 

\begin{equation}
 [A] = [U] \, [\Sigma]\,[V]^T,    
\label{eq:svd}
\end{equation}

where $U$ (shape $n \times r$) and $V$ (shape $r \times m$) are orthonormal matrices and $\Sigma$ a diagonal matrix containing the $r$ singular values of $A$. Note that generally all references use the symbols $U$ and $V$ to present SVD, which are also used in this work for consistency. This choice may, however, be confusing as the symbol $U$ is also used for the matrices in which the input signals are recast (Equations \ref{eq:inputugenral}, \ref{eq:inputustep}) but we do so because most of the references on the topic use the same symbol.

In the case Volterra kernels are projected onto the subspace defined by $l$ Laguerre polynomials:

\begin{equation}
    \{h\}  = [B]\,\{ \theta \}
\end{equation}

where $[B]$ is a $m\times l$ matrix. Eq.~\ref{eq:convolution_1} is re-written as:

\begin{equation}
    \{ y \} = [U]\, [B]\,\{ \theta \},  
    \label{eq:convolution_L1}
\end{equation}

and also solved using SVD to obtain the pseudo-inverse ($( [U]\, [B] )^+$) of the $n \times l$ matrix $ [U]\, [B]$:

\begin{equation}
    \{ \theta \} = ( [U]\, [B] )^+ \, \{ y \},
\end{equation}

\subsubsection{Identification of Higher-order Volterra kernels for a SISO System}

Following Ref.~\cite{levin2022convolution}, a two-step approach is exploited. It is not applicable in general and only justified if the linear part of the response  dominates over the nonlinear contribution.   

The approach exploits a number of input signals with identical shape but increasing magnitude ($u_A$, $u_B$, \dots, $u_N$) and the corresponding CFD responses, $y_A$, $y_B$, \dots, $y_N$. Attention must be taken to choose the magnitude of $u_A$ such that the linear response alone provides the necessary accuracy. 
Linear kernels may then be identified using Eq.~\ref{eq:linearsolution_1} from the input signal of magnitude $\alpha_A$: 
\begin{equation}
    \{ h_1 \} = [U_A]^{-1} \{y_A\},
    \label{eq:linearsolution_1A}
\end{equation}
where the matrix $[U_A]$ corresponds to $[U]$ in Eq.~\ref{eq:inputugenral}.
To a minimum, one requires at least one response with magnitude $\alpha_B > \alpha_A$.
Then, one can assume that $y_B$ can be approximated with a second order Volterra series: 
\begin{equation}
    \{y_B\}  =  [U_B] \{h_1\} + [U_{B2}] \{h_2\}, 
    \label{eq:volterra_B2}
\end{equation}
where $[U_B]$ is build like the matrix $[U]$ in Eq.~\ref{eq:inputugenral} and $[U_{B2}]$ contains the square of the terms of the input signal, in the same form of $[U_B]$: 
\begin{equation}
    [U_{B2}] = \left[
    \begin{array}{ccccc}
       u(0)^2  &  \, & \, & \,& \, \\
       u(1)^2 &  u(0)   & \, & \, & \, \\ 
       u(2)^2 &  u(1)^2   & u(0)^2, & \, & \, \\ 
       \dots & & & & \\ 
       u(m)^2 & u(m-1)^2 & u(m-2)^2 & \dots & u(0)^2  \\ 
       \dots & & & & \\ 
       u(n)^2 & u(n-1)^2 & u(n-2)^2 & \dots & u(n-m)^2  \\ 
    \end{array}
    \right],
\label{eq:inputugenral2}    
\end{equation}
In particular, when the input signal is a step function, $[U_{B2}]$ reads: 

\begin{equation}
    [U_{B2}] = \alpha^2_B \, \left[
    \begin{array}{ccccc}
       1  &  0 & 0 & \dots & 0 \\
       1  &  1 & 0 & \dots & 0 \\ 
       \dots & & & & \\ 
       1 & 1 & 1 & 1 & 1 \\ 
    \end{array}
    \right].
\end{equation}
Eq.~\ref{eq:volterra_B2} shows that the second order kernels, $h_2$, can be determined to maximise the accuracy of  $y_B$ reconstruction:
\begin{equation}
     [U_{B2}] \{h_2\} =  \{y_B\} - [U_B] \{h_1\}, 
    \label{eq:convolution_2}
\end{equation}

leading to the solution for $h_2$: 

\begin{equation}
    \{h_2\} = [U_{B2}]^+ \, (\{y_B\} - [U_B] \{h_1\}). 
\end{equation}

If more than two signals are available, higher order kernels can be identified. For instance, if two signals $U_B$ and $U_C$ and their responses are known, a third order series:

\begin{equation}
 \{ y_B\}  =  [U] \{h_1\} + [U_{2}] \{h_2\} + [U_{3}] \{h_3\},   
\end{equation}
where $ [U_{3}]$ is analogous to $[U_{2}]$ but with the input terms raised to the third power. For the signals $B$ and $C$: 
\begin{equation}
    \begin{array}{l}
      \{ y_B\}  =  [U_B] \{h_1\} + [U_{B2}] \{h_2\} + [U_{B3}] \{h_3\} \\
      \{ y_C\}  =  [U_C] \{h_1\} + [U_{C2}] \{h_2\} + [U_{C3}] \{h_3\},
    \end{array}
\label{eq:identification_23orders}
\end{equation}

$h_2$ and $h_3$ can be identified  from Eq.~\ref{eq:identification_23orders}:

\begin{equation}
    \begin{array}{l}
      \, [U_{B2}] \{h_2\} + [U_{B3}] \{h_3\} = \{ y_B\}  -  [U_B] \{h_1\} \\
      \, [U_{C2}] \{h_2\} + [U_{C3}] \{h_3\} = \{ y_C\}  -  [U_C] \{h_1\},
    \end{array}
\end{equation}
simultaneously solving for $h_2$ and $h_3$: 
\begin{equation}
    \left\{
    \begin{array}{l}
      \{h_2\}    \\
      \{h_3\}   
    \end{array}
    \right\}
    =
    \left[
    \begin{array}{cc}
      U_{B2}   &  U_{B3}\\
      U_{C2}   &  U_{C3}
    \end{array}
    \right]^+ \, 
    \left\{
    \begin{array}{l}
       \{ y_B\}  -  [U_B] \{h_1\}    \\
       \{ y_C\}  -  [U_C] \{h_1\}   
    \end{array}
    \right\}.
\label{eq:identification_23_sim}
\end{equation}

As pointed out in~\cite{levin2022convolution}, Eq.~\ref{eq:identification_23_sim} is not the only possible strategy. Results published in this paper were obtained with second order kernels only, in order to keep  within limits the computational demand for the generation of responses. However, different tests with  higher-order series were carried out for few parameter samples, showing the effectiveness Eq.~\ref{eq:identification_23_sim} and its superiority with respect to a multi-step approach ($h_2$ from response $y_B$, then $h_3$ from response $U_c$ etc).

\subsection{Artificial Neural Network}

The first method used to approximate the Volterra kernels in the \(M-\alpha_0\) design space is the Artificial Neural Network (ANN), a common supervised learning tool. A fully--connected neural network (FCNN) architecture is employed, where each neuron is connected to all neurons in the next layer. The output of each neuron is calculated as:

\begin{equation}  \label{eq:FCNN}
y_{l,n} = f_l( w_{l,n} \cdot x_l + b_{l,n} )
\end{equation}

where \(l\) is the layer number, \(n\) is the neuron number, \(w_{l,n}\) the weights, and \(b_{l,n}\) the bias. The function \(f_l()\) is the nonlinear activation function applied to the weighted sum.

In this study, the FCNN predicts both the first-order and second-order Volterra kernels, which are crucial for accurately modeling the system response. These kernels are mathematically represented by Equations~\ref{eq:1st_order_volterra}-\ref{eq:2nd_order_volterra}, which define the linear and nonlinear components of the system. The interpolation of these kernels is necessary because the Volterra series, as expressed in Equation~\ref{general_volterra_series}, relies on these kernels to capture the unsteady aerodynamic behavior across different $M$ and \(\alpha_0\). 
The predicted kernels are then used in the convolution operation described by Equation~\ref{eq:volterra_B2}, which allows the reconstruction of the system response to various inputs. This process is central to applying the Volterra series framework effectively, particularly in scenarios involving complex, nonlinear aerodynamic phenomena.

The FCNN architecture consists of an input layer for \(M\), \(\alpha_0\), and static values of the lift coefficient $C_{L0}$ and pitching moment coefficient $C_{M0}$, followed by hidden layers and an output layer predicting the Volterra kernels. To capture the different dynamics of the system, two separate models were trained: one for the linear kernels and another for the nonlinear kernels. This distinction is necessary because linear and nonlinear kernels represent fundamentally different aerodynamic behaviors—linear kernels dominate small perturbations, while nonlinear kernels capture higher-order effects, especially under strong nonlinear conditions. By training distinct models, we ensure higher accuracy for each type of kernel.

The Mean Squared Error (MSE) loss function was used to evaluate the model performance, quantifying the difference between predicted and true output. The Adam optimization algorithm was utilized during the back-propagation phase to optimize the neural network weights and minimize the MSE loss function.

To enhance the predictive capability of the model, selecting an appropriate set of hyperparameters is crucial, and requires an optimization algorithm that can explore the vast design space. We also used a Bayesian optimization approach to optimize the hyperparameters of the network. This automated technique treats the entire hyperparameter space as a black box, eliminating the need for manual studies or conventional sensitivity analyses, and allows for determination of the impact of hyperparameters on the results. Table~\ref{tab_Hyperparameters_design_space} shows the hyperparameters design space in terms of values and step size that each variable may assume. For a detailed description of the formulation, the reader is referred to Immordino et al.~\cite{immo2023ROM}.

\begin{table}[!htb]
\centering
\begin{tabular}{l c c }
\hline 
\hline 
\textbf{Hyperparameter} & \textbf{Value} & \textbf{Step size} \\ 
\hline
Learning rate   & $10^{-4}$ to $10^{-6}$     &  $5\cdot10^{-6}$  \\  
Number of Hidden Layers   & 1 to 8   &  1 \\  
Number of Neurons per Hidden Layer  & 4 to 204   & 8 \\  
Batch size   & 1 to 8  &  1   \\  
Activation function   & TanH, ReLu, PReLu  & /  \\  

\hline
\hline 
\end{tabular}
\caption{Design space of the hyperparameters used in the Bayesian optimization.}
\label{tab_Hyperparameters_design_space}
\end{table}

\subsection{Gaussian Process Regression}

The second technique analyzed for reconstructing the Volterra kernels is the Gaussian Process Regression (GPR). It is a statistical technique for regression analysis that is non-parametric and Bayesian in nature. Unlike conventional linear regression method, GPR does not rely on assumptions about the relationship between input and output variables and models it as a probabilistic distribution rather than a deterministic equation. The GPR model consists of a mean function, a covariance function (kernel), and a collection of hyperparameters that regulate the shape and scale of the covariance function. When provided with a new input, the GPR model returns a distribution for the corresponding output, which facilitates probabilistic forecasts and uncertainty evaluations. 

As with the FCNN model, the GPR is fed with inputs consisting of $M$, $\alpha_0$, and the static values of $C_{L0}$ and $C_{M0}$, and outputs the Volterra kernels. Two different models were trained: one for the linear kernels and another for the nonlinear ones.

The Gaussian process is defined by a mean function \(m(x)\) and a covariance function \(k(x, x')\), which represent the mean and covariance of the outputs for given inputs:

\begin{equation}
\mu(x) = E[f(x)]
\end{equation}

\begin{equation}
K(x, x') = Cov[f(x), f(x')]
\end{equation}

where $E$ is the expected value and $Cov$ is the covariance. The mean function can be any deterministic function, while the covariance function can be any positive definite function. The mean function $\mu(x)$ can be set to zero without loss of generality, since it just shifts the prediction for $f(x)$ by a constant. The covariance function $k(x, x')$ encodes the similarity between the inputs $x$ and $x'$. It is usually referred to as a kernel, and different types of kernels can be used to model different types of relationships between inputs and outputs. In our case, a linear combination of radial basis function (RBF) and Matern 5/2 kernel led to the most accurate results. RBF kernel can be expressed as follow:

\begin{equation}
k_{RBF}(x,x') = \sigma_f^2 \exp\left(-\frac{1}{2l^2} |x-x'|^2\right)
\end{equation}

where $\sigma_f^2$ is the signal variance, $l$ is the length scale parameter and $||x-x'||^2$ is the squared Euclidean distance between the inputs $x$ and $x'$. The length scale 
parameter determines the smoothness of the function and controls the influence of each data point on the prediction.

 The Matern 5/2 covariance function is defined as:

\begin{equation}
k_{Matern 5/2}(x, x') = \sigma_f^2 \left(1+\frac{\nu|x-x'|}{l}+\frac{(\nu|x-x'|)^2}{3l^2}\right)\exp\left(-\frac{\nu|x-x'|}{l}\right)
\end{equation}

where $\nu$ is the smoothness parameter, equal to $\sqrt{5}$ in this case. 

To optimize the hyperparameters of the covariance functions and maximize the likelihood of the observed data, we used the Limited-memory BFGS with bounds (L-BFGS-B) quasi-Newton optimization algorithm.

\section{Synthetic Aerodynamics}\label{sec:synthetic_aero}

To demonstrate the validity and feasibility of the concept, we utilize a system of differential equations specifically designed to replicate the indicial responses of real wings and aerofoils. This approach allows for the efficient generation of data while providing control over the system's nonlinearity and noise levels. The differential equations mimic the state-space formulation proposed by Leishman~\cite{leishman1993indicial}:



\begin{equation} \label{eq:aero_lags}
\begin{split}
\cfrac{dx_1}{d\tau} & = -\cfrac{2V}{c} b_1 x_1 \\
\cfrac{dx_2}{d\tau} & = -\cfrac{2V}{c} b_2 x_2 \\
\cfrac{dx_3}{d\tau} & = -\cfrac{2V}{c} b_3 x_3, \\
\end{split}
\end{equation}
where $x_1$, $x_2$, $x_3$ play the role of the aerodynamic lag states and $\tau$ is the reduced time. 
We introduce a first, linear, expression of the  lift coefficient:

\begin{equation} \label{eq:linear_cl}
C^l_L(\tau) = C_{L/\alpha} \left( A_1 b_1 x_1(\tau)  + A_2 b_2 x_2(\tau) + A_3 b_3 x_3(\tau)\right),
\end{equation}
where the coefficients \(C_{L/\alpha} = 2 \pi\), \(A_1 = 0.670\), \(A_2 = 0.330\), \(b_1 = 0.30\), and \(b_2 = 0.0455\) are modified from Jones approximation of the Wagner function. We introduced a dependence of the system coefficients on two parameters: \(M\) and \(\alpha_0\). The lift curve slope is influenced by \(M\) through the Prandtl-Glauert factor, while the coefficient \(A_3\) depends nonlinearly on both \(M\) and \(\alpha_0\), ranging from 0 to \(-0.15\). The value \(b_3 = 0.15\) was selected to approximate a dynamic stall response.

We also introduce a second, nonlinear expression of the lift coefficient, \(C^{nl}_L\):

\begin{equation} \label{eq:nonlinear_cl}
    C^{nl}_L(\tau) = C_{l/\alpha} \left( A_1 b_1 x_1(\tau)  + A_2 b_2 x_2(\tau) + A_3 b_3 x_3(\tau) - C_{nl} x_2(\tau) x_3(\tau) \right),
\end{equation}

The nonlinearity term \(C_{nl} = -0.35\) was selected to mimic the qualitative behavior observed in transonic aerofoil responses obtained from CFD simulations. To account for variability, white noise was introduced to all signals. A summary of our analysis is provided in Algorithm~\ref{alg:algorithm1}.


\begin{algorithm}[hbt!]
\caption{Procedure for reconstructing the system response using Volterra series and ML}
\label{alg:linear}
\begin{algorithmic}[1]
\STATE Generate $n$ samples within the parametric space defined by $M$ and $\alpha_0$.
\STATE Calculate the response for each sample to a step input signal with an amplitude of one degree using Equations \ref{eq:aero_lags} and \ref{eq:linear_cl}.
\STATE Compute the linear Volterra kernels based on the responses.
\STATE Calculate additional responses to a step input signal with an amplitude of two degrees using Equations \ref{eq:aero_lags} and \ref{eq:nonlinear_cl} to account for nonlinear effects.
\STATE Compute the nonlinear Volterra kernels based on the additional responses.
\STATE Split the $n$ samples into three subsets: training ($n_1$), test ($n_2$), and validation ($n_3$).
\STATE Use the training and test subsets ($n_1$ and $n_2$ samples) to train the GPR and FCNN models.
\STATE Predict the Volterra kernels for the validation subset ($n_3$ samples) using the trained GPR and FCNN models.
\STATE Reconstruct the system response to a sinusoidal input in the validation subset using the predicted Volterra kernels and compare these results with the exact responses from the differential equations.

\end{algorithmic}
\label{alg:algorithm1}
\end{algorithm}

\begin{figure} [!htb] 
    \centering
    \includegraphics[trim=0 0 0 0, clip, width=0.98\textwidth]{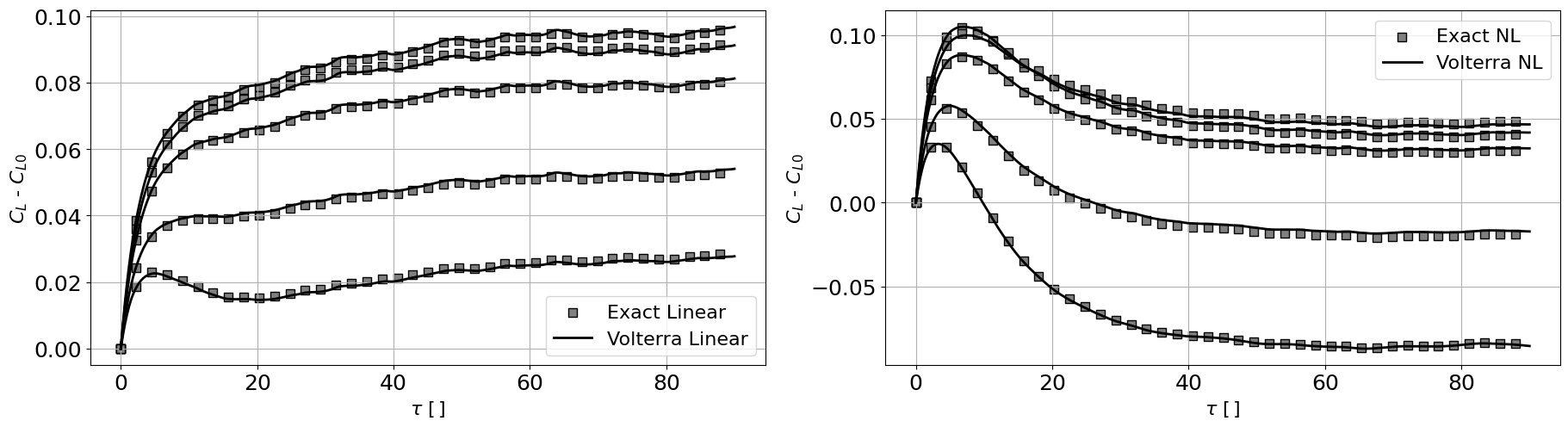}
    \caption{Examples of linear (left) and nonlinear (right) synthetic responses under a step input. The squares indicate the exact responses, the continuous lines denote the responses reconstructed with Volterra series. 
    }
    \label{fig:testset_stepresponses}
\end{figure}

We defined the parameter space for \(M\) ranging from 0.40 to 0.85 and \(\alpha_0\) between -2 and 8 degrees. Using Latin hypercube sampling (LHS), we generated 70 data points for both variables. These were divided into $n_1 = 45$ for training, $n_2 = 15$ for test, and $n_3 = 10$ for validation. 
Figure~\ref{fig:testset_stepresponses} illustrates a few  examples of the synthetic time responses for both the linear and nonlinear systems under a step input. The right panel presents the nonlinear system response and its corresponding reconstruction. The linear response follows a smooth, predictable trajectory, which is accurately captured by the first-order Volterra kernel reconstruction. The close alignment of the continuous line with the square markers demonstrates that the first-order Volterra kernels can effectively replicate the exact system behavior, as expected for a linear system. In contrast, the nonlinear system exhibits a more complex response. The continuous line representing the second-order Volterra kernel reconstruction shows small deviations from the true response (square markers), particularly in the dynamic stall region. This highlights the challenge of capturing nonlinear effects using a second-order Volterra series. Despite these differences, the reconstructed response still provides a reasonably accurate approximation of the true behavior, especially in regions of moderate nonlinearity.

\begin{figure}[!htb]
    \begin{center}
    
    \subfigure[{Linear responses to a sinusoidal pitch input of $\pm 1$ [deg]}] 
    {\label{linear_responses}
      \includegraphics[trim=0 0 0 0, clip, width=0.48\textwidth]{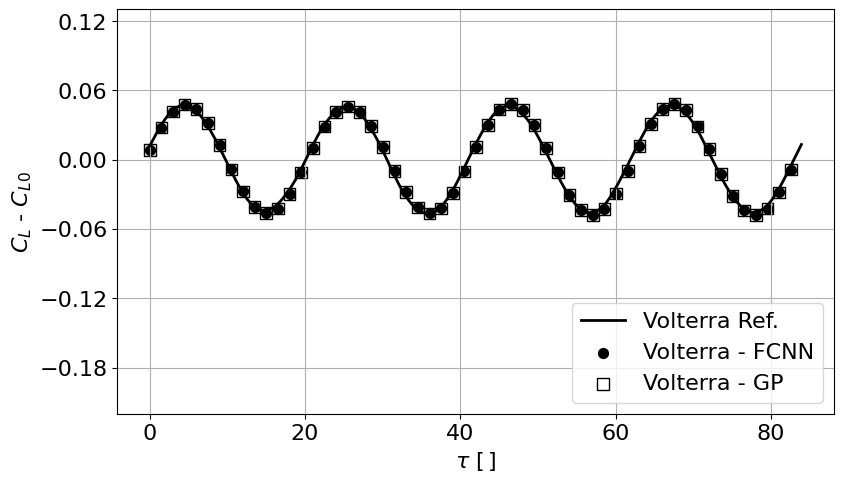}
      \includegraphics[trim=0 0 0 0, clip, width=0.48\textwidth]{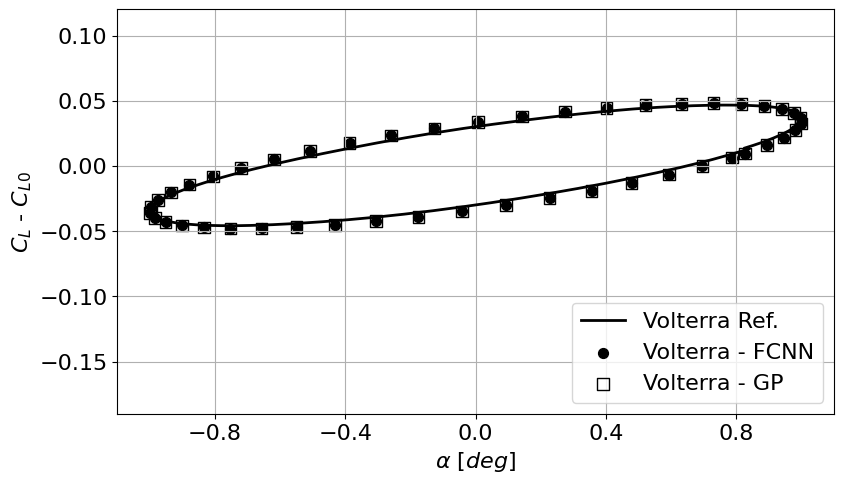}} \\

    \subfigure[{Nonlinear responses  to a sinusoidal pitch input of $\pm 2$ [deg]}] 
    {\label{nonlinear_responses}
      \includegraphics[trim=0 0 0 0, clip, width=0.48\textwidth]{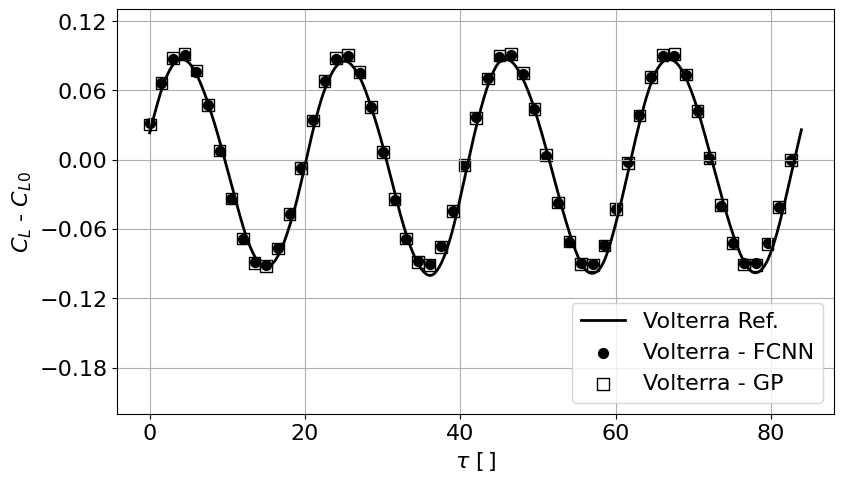}
      \includegraphics[trim=0 0 0 0, clip, width=0.48\textwidth]{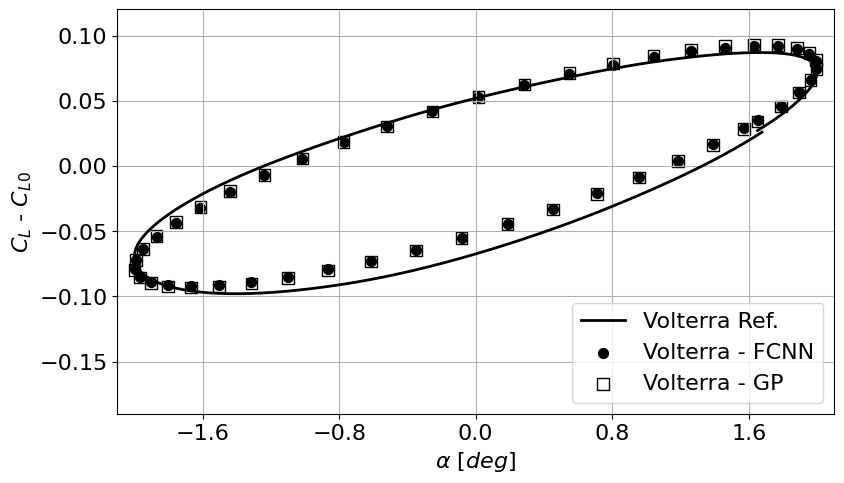}}

    \caption{Examples of linear (top) and nonlinear (bottom) synthetic responses under a sinusoidal input at reduced frequency $k = 0.3$. Continuous lines indicate the reference Volterra series derived from the exact solutions, while circles and squares denote the reconstructed responses using kernels predicted by FCNN and GPR, respectively.}
    \label{fig:synthetic_responses}
    
    \end{center}
\end{figure}

After having determined both linear and nonlinear Volterra kernels for the $n_1$ training samples and $n_2$ test samples within the $M-\alpha_0$ design space, we trained the GPR and FCNN algorithms. The trained models were then used to reconstruct the Volterra kernels in the validation set and calculate the $n_3$ responses to a sinusoidal input with a reduced frequency \(k = 0.3\) and two different amplitudes (last line in Algorithm \ref{alg:algorithm1}). We used the smaller amplitude signal as input to the linear model (Eq.~\ref{eq:linear_cl}) and  the larger one as input to the nonlinear model (Eq.~\ref{eq:nonlinear_cl}).  
Figure~\ref{fig:synthetic_responses} shows both the time response (left) and the phase plane (right). It is worthwhile mentioning that while first-order kernels can exactly reconstruct linear system responses via convolution integrals, responses reconstructed using both first- and second-order kernels are not exact. Both ML methods accurately predict the linear response, while exhibiting similar minor deviations when applied to the nonlinear case. These results give us confidence in the potential of the method for more complex applications, as explored in the later test cases in this document.

\section{Two-dimensional Test Case}\label{sec_naca0012}

The second test case is for a two-dimensional aerofoil section based on the NACA0012 profile. The choice was motivated by the presence of highly nonlinear phenomena in the transonic regime, characterized by the formation of intense shock waves on both the upper and lower surface, often associated with induced separations of the boundary layer.

Computational Fluid Dynamics (CFD) indicial step--responses were generated through the Unsteady--RANS (URANS) formulation, employing SU2 v7.5.0—a cell-centered  finite volume software suite~\cite{economon2016su2}, with the one--equation Spalart-Allmaras model; the negative turbulence production option was always active.
To accelerate the convergence of CFD simulations, we adopted a $1v$ multigrid scheme. For discretizing convective flows, we used the JST central scheme with artificial dissipation, while flow variable gradients were calculated using the Green Gauss method. Linear solving was accomplished with the biconjugate gradient stabilization method, complemented by ILU preconditioning. URANS simulations were restarted from the steady--state solution. The nondimensional timestep employed was $\Delta \tau=0.15$, and the total nondimensional time of the simulations was set to $\tau=113.6$, allowing for the complete development of the flow during the step motion.

\begin{figure} [!htb] 
    \centering
    \includegraphics[trim=1 1 1 1, clip, width=0.48\textwidth]{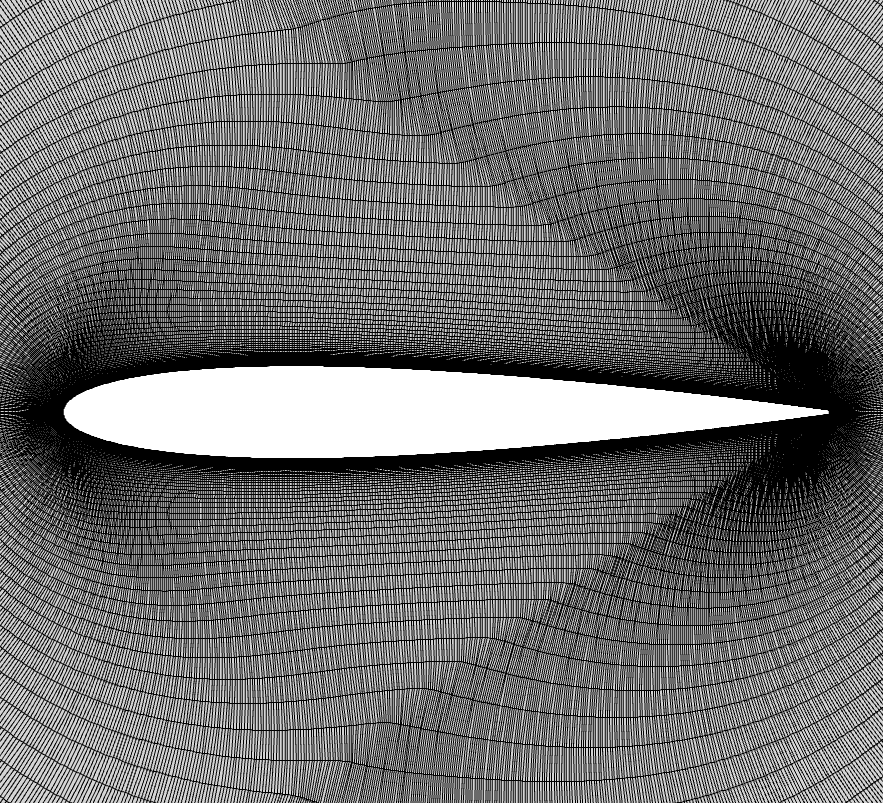}
    \caption{Impression of the NACA0012 CFD grid.}
    \label{fig_naca_mesh}
\end{figure}

A structured O--mesh with 177,112 elements was generated. To maintain a consistent boundary layer resolution and adequately capture shock waves, we adopted a $y^+ = 1$. The computational domain spans 100 chord lengths from the solid wall to the farfield boundary. For a visual representation of the mesh, refer to Figure~\ref{fig_naca_mesh}.

\begin{figure} [!htb] 
    \centering
    \includegraphics[trim=0 0 0 0, clip, width=0.49\textwidth]{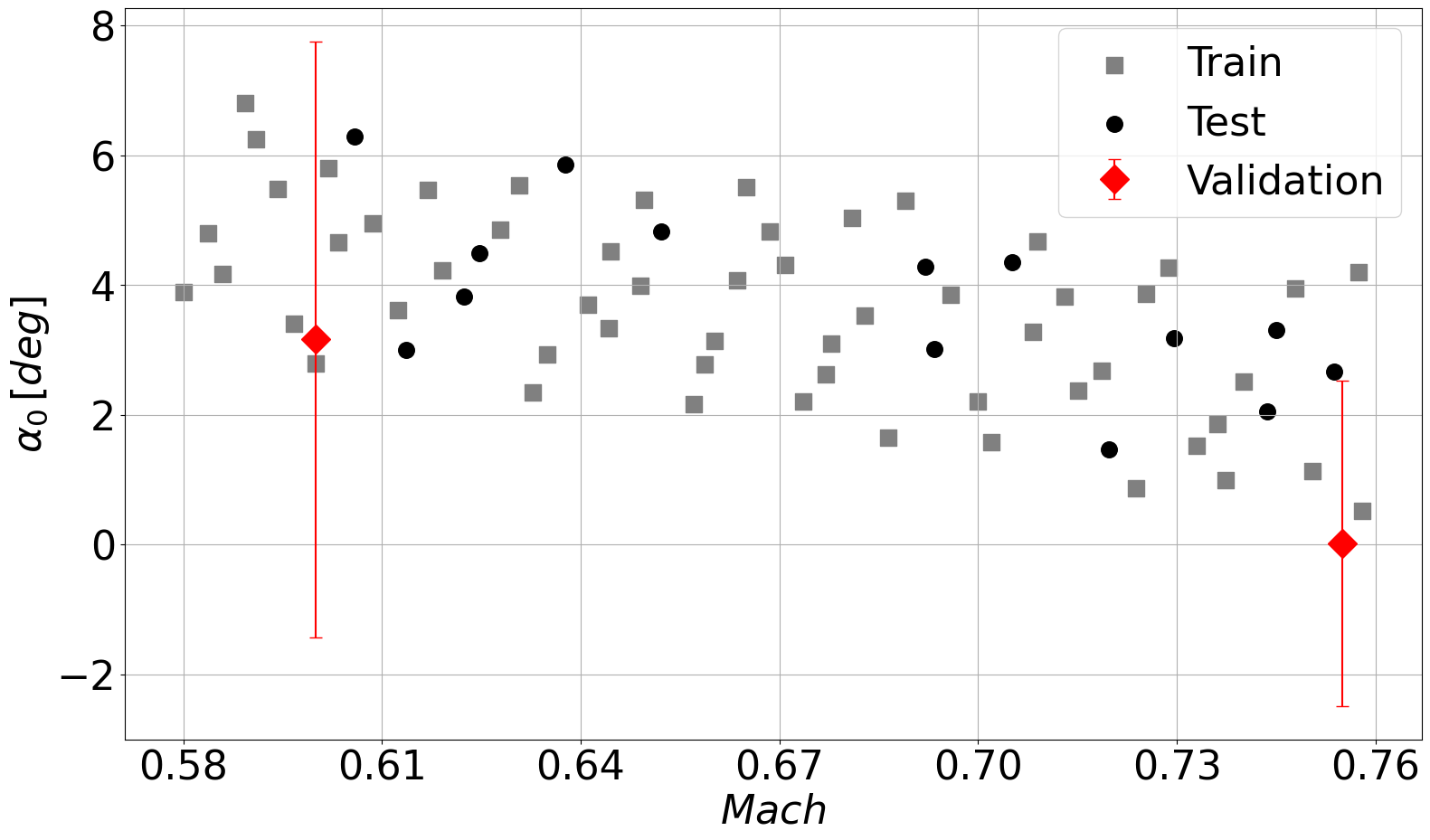}
    \caption{Samples for $M$ and $\alpha_0$ for NACA0012 test case. Red markers denote flight conditions of CT2 and CT2 experiments~\cite{landon1982naca}. }

    \label{fig_sampled_points_2D}
\end{figure}

We defined a parameter space for $M$ (0.58 to 0.76) and $\alpha_0$ (0 to 7 deg) to capture the physics associated with the shock-boundary layer interaction. Using LHS, we generated 70 data points for both variables (see Figure~\ref{fig_sampled_points_2D}). The data was split into 80\% for training and 20\% for test to evaluate the model performance. The excitation signal for each of the 70 step responses consists of an indicial plunge response generated with vertical velocities of $-1 \, m/s$ and $-2 \, m/s$, enabling the identification of both linear and nonlinear dynamics. Red markers in Figure~\ref{fig_sampled_points_2D} denote flight conditions of CT2 and CT5 experiments~\cite{landon1982naca} with the bands indicating their variations in pitch during the harmonic forced motions. These two signals are used as validation points, where the prediction involves harmonic motion, with both mean and amplitude being critical variables. We may claim that this approach tests the model predictive capability beyond the bounds of the training region.

Similar to the previous test case, the initial step involved identifying both the linear and nonlinear Volterra kernels for the training and test samples within the \(M-\alpha_0\) design space. We then employed the GPR and FCNN algorithms to predict the kernel coefficients at the points in the validation set, and to reconstruct the harmonic signal associated with the AGARD CT2 and CT5 cases~\cite{landon1982naca}.

\begin{figure}[!htb]
\begin{center}
\subfigure[{CT2: $M= 0.60$ -- $f=50.32 \, Hz$ -- $\bar{\alpha}= 3.16$ $[deg]$ -- $\alpha_A=4.59$ [deg]}] {\label{harmonic_signal_M060}  \includegraphics[trim=0cm 0cm 0cm 1cm, clip, width=0.99\textwidth]{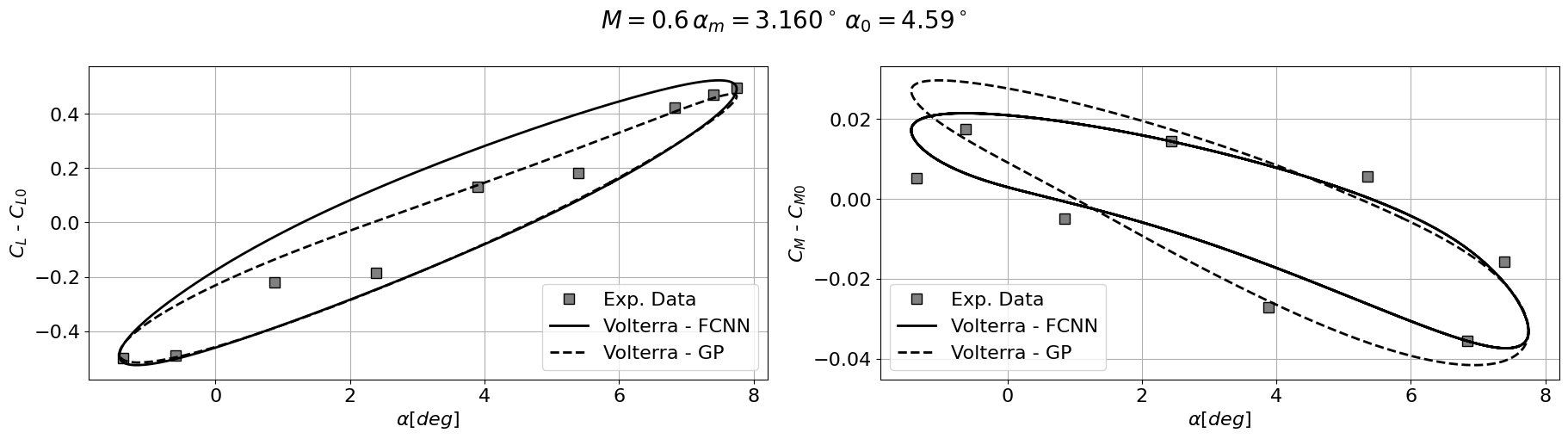}} \\ 
\subfigure[{CT5: $M= 0.755$ -- $f=62.5 \, Hz$ -- $\bar{\alpha}= 0.016$ $[deg]$ -- $\alpha_A=2.51$ [deg]}] {\label{harmonic_signal_M0755}
  \includegraphics[trim=0cm 0cm 0cm 1cm, clip, width=0.99\textwidth]{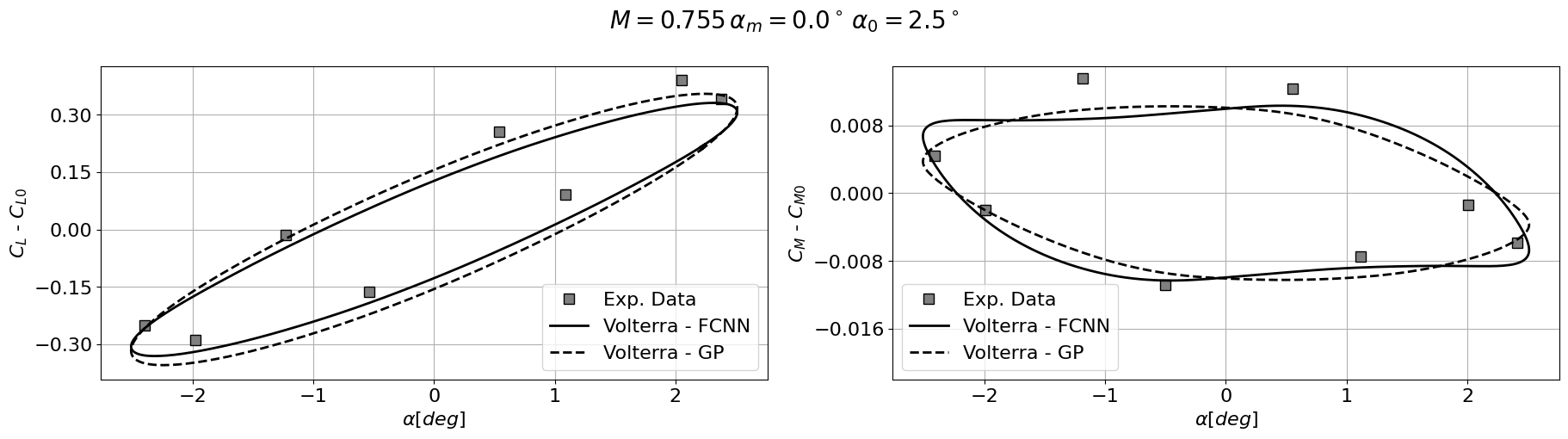}}
    \caption{Harmonic signal reconstruction with Volterra ROM for the AGARD CT2 and CT5 cases~\cite{landon1982naca}. Experimental data are represented by square markers, while continuous and dashed lines denote the reconstructed responses using nonlinear Volterra kernels predicted by FCNN and GPR, respectively.}
    \label{fig_harmonic_signal_reconstruction_2D}
    \end{center}
\end{figure}

For the CT2 case, the parameters are \(M = 0.60\), frequency \(f = 50.32\, \text{Hz}\), mean angle of attack \(\bar{\alpha} = 3.16^\circ\), and amplitude of pitch oscillation \(\alpha_A = 4.59^\circ\). Figure~\ref{harmonic_signal_M060} shows the results for this case. The hysteresis loops for lift coefficient \(C_L\) and pitching moment coefficient \(C_M\) are presented, comparing the experimental data to the nonlinear Volterra reconstructions using FCNN (continuous line) and GPR (dashed line). Both ML models provide a good reconstruction of the \(C_L\) hysteresis. However, for the \(C_M\) hysteresis, a slight discrepancy is observed, likely due to the significant pitch variation, which challenges the accuracy of the Volterra series. Despite this, FCNN captures the dynamic behavior of \(C_M\) more effectively than GPR. 

For the CT5 case, the parameters are \(M = 0.755\), frequency \(f = 62.5\, \text{Hz}\), mean angle of attack \(\bar{\alpha} = 0.016^\circ\), and amplitude of pitch oscillation \(\alpha_A = 2.51^\circ\). Figure~\ref{harmonic_signal_M0755} shows the corresponding hysteresis loops for \(C_L\) and \(C_M\). In this case, FCNN outperforms GPR for both aerodynamic coefficients. The \(C_L\) hysteresis is well-captured by both models, while \(C_M\) hysteresis shows some deviations, with FCNN providing a more accurate fit.

Overall, the nonlinear Volterra reconstruction of the aerodynamic coefficients was reasonably good in both cases, particularly for \(C_L\) hysteresis, aligning with CFD results from Yao et al.~\cite{yao2016nonlinear}. However, minor discrepancies were found in \(C_M\), especially in the CT2 case, likely due to the wide pitch variation during the oscillation. The significant variations in the kernels coefficients between training points make this a highly challenging test case for the proposed methodology. Exploring the use of higher--order kernels could potentially mitigate this issue, although it would dramatically increase the computational demands.

\section{Three-dimensional Test Case}\label{sec_bscw}

To further increase the complexity of the model geometry and flow features, the third test case is the Benchmark Super Critical Wing (BSCW). It is a rigid semi--span wing with a rectangular planform and a supercritical aerofoil shape from the AIAA Aeroelastic Prediction Workshop~\footnote{\url{https://nescacademy.nasa.gov/workshops/AePW3/public/wg/highangle}}. The wing is mounted on a flexible system with two degrees of freedom, allowing movement in both pitch and plunge. However, in this study, we focus exclusively on pitch responses. The BSCW has been specifically designed for flutter analysis 
in the presence of shock-boundary layer interaction. 

\subsection{Reference Dataset} \label{sec_cfd_data}

\begin{figure} [!htb] 
    \centering
    \includegraphics[trim=1 1 1 1, clip, width=0.48\textwidth]{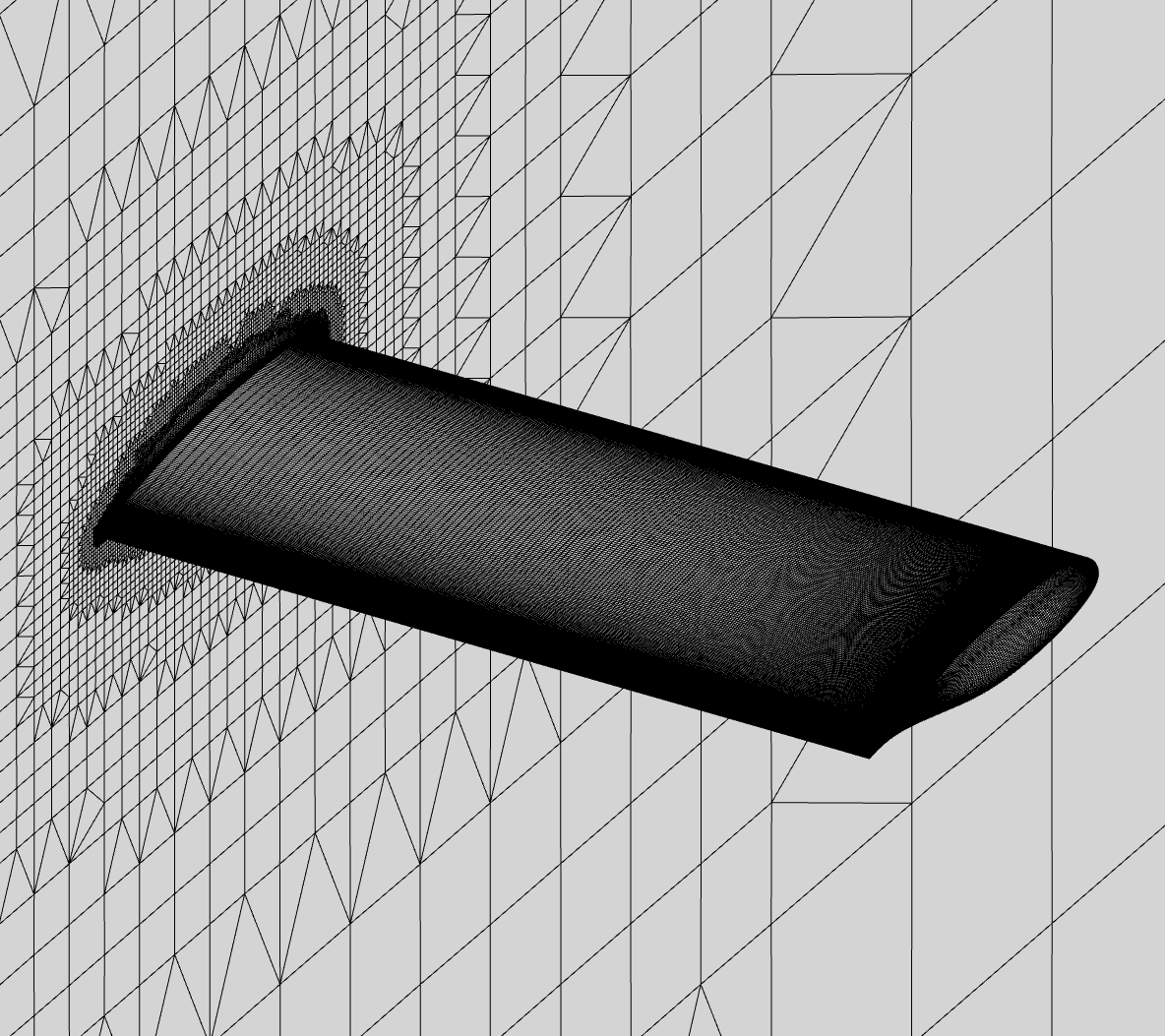}
    \caption{Impression of the BSCW CFD grid.}
    \label{fig_bscw_mesh_1}
\end{figure}

Also in this test case, we generated CFD indicial step--responses using URANS formulation with SU2 v7.5.0. The nondimensional time step employed was $\Delta \tau=0.029$. The total nondimensional time of the simulations was set to $\tau=27.2$ to ensure full development of the flow. A mixed-type grid with $15.6 \cdot10^6$ elements and 130,816 surface elements was generated, structured on the wing surface and in the first layers of the boundary layer, voxel in the rest of the computational domain. A $y^+ = 1$ is adopted, after a preliminary mesh convergence study that ensured an adequate resolution of the boundary layer and shock wave. The computational domain extends 100 chords from the solid wall to the farfield. An impression of the grid can be obtained from Figure~\ref{fig_bscw_mesh_1}. 

\begin{figure} [!htb] 
    \centering
    \includegraphics[trim=0 0 0 0, clip, width=0.995\textwidth]{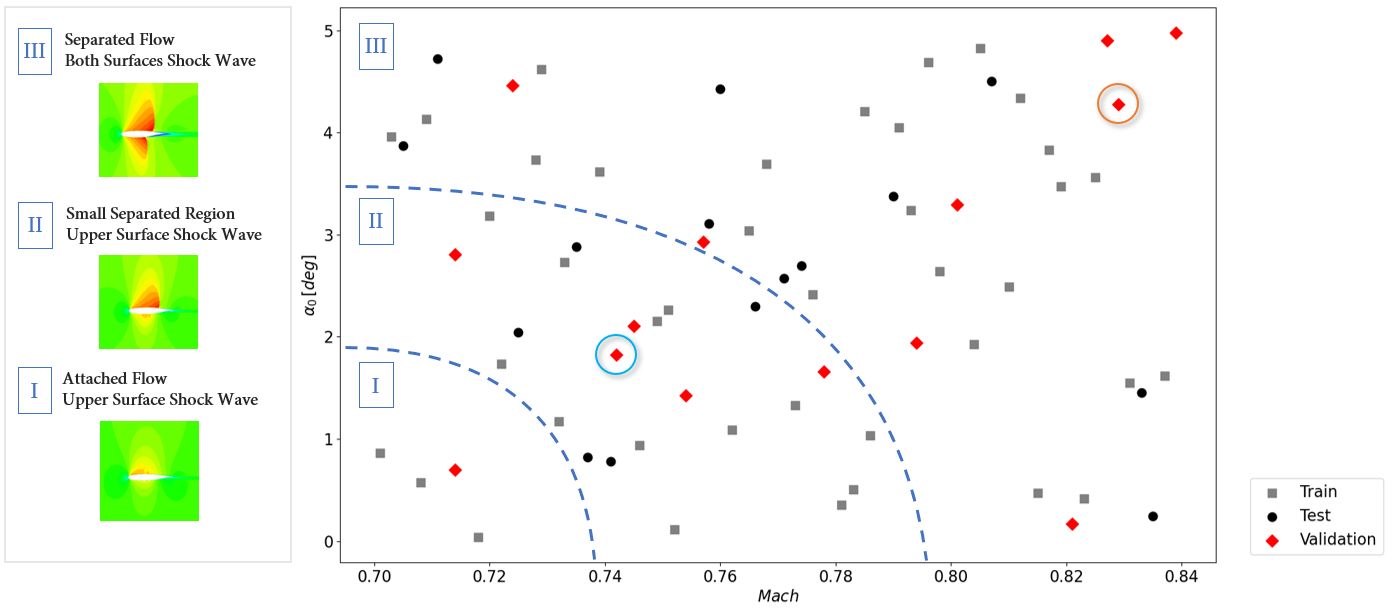}
    \caption{Training, test and validation samples for $M$ and $\alpha_0$. Three regions define the increasing complexity of the physics phenomena to be captured. Blue and orange circles indicate the conditions analyzed in this section.}
    \label{fig_sampled_points_1}
\end{figure}

\begin{figure} [!htb] 
    \centering
    \includegraphics[trim=1.5 1.5 1.5 1.5, clip, width=0.98\textwidth]{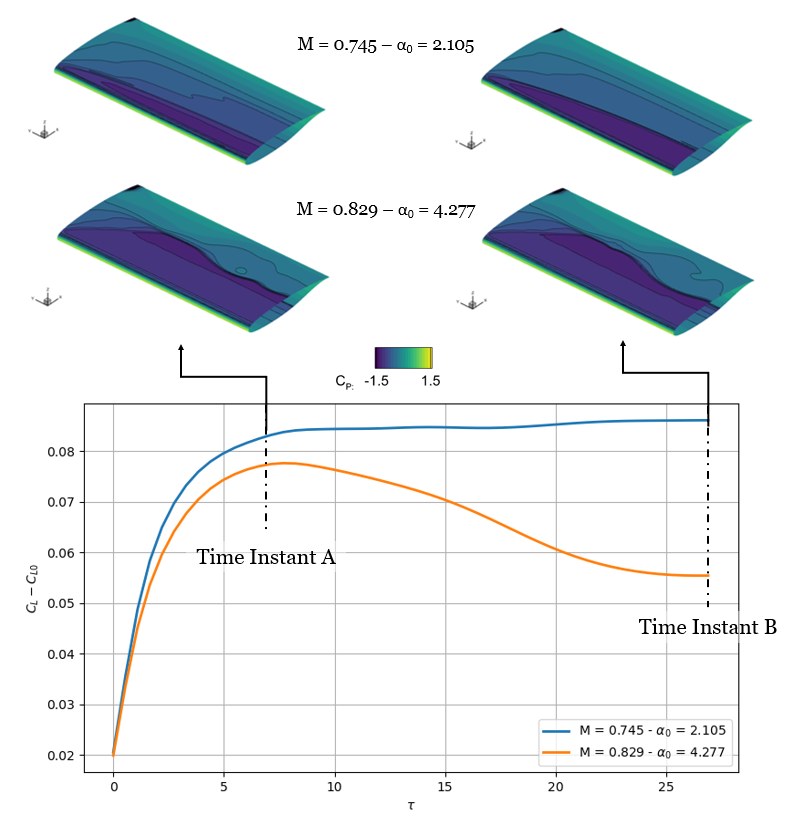}
    \caption{Lift coefficient responses at two different flight conditions with pitch input angle of 1 $[deg]$. The steady--state values were subtracted. Pressure coefficient contours are also added at $\tau=7$ denoted as $Time \, Instant \, A$, and $\tau=30$ denoted as $Time \, Instant \, B$.}
    \label{fig_volterra_step_linear_nonlinear}
\end{figure}

Following the same approach as in the other cases, the two independent parameters for the models are $\alpha_0$ and $M$ with ranges of $[0,5]$ degrees and $[0.70,0.84]$, respectively. These ranges were selected for consistency with the test cases in the Second and Third  Aeroelastic Prediction Workshops. The necessary $n$ samples for $\alpha_0$ and $M$ are defined through LHS with a total of 70 points (Figure~\ref{fig_sampled_points_1}). Of these, 60\% are selected for training, 20\% for testing, and the remaining 20\% for validation. The three regions marked in Figure~\ref{fig_sampled_points_1} highlight the increasing complexity of the physics phenomena, in terms of boundary layer separation and shock wave formation on both sides of the wing. Region I, at lower $M$ and $\alpha_0$, represents conditions where the flow remains mostly attached to the surface, and shock waves are either weak or absent, resulting in minimal boundary layer separation. Region II corresponds to moderate $M$ and $\alpha_0$, where the onset of shock-induced boundary layer separation begins to occur, particularly on the upper surface of the wing, leading to more complex aerodynamic behavior. Region III, at higher $M$ and $\alpha_0$, captures the most challenging aerodynamic conditions, with strong shock waves forming on both the upper and lower surfaces of the wing, resulting in extensive boundary layer separation and significantly nonlinear flow characteristics. This progressive increase in aerodynamic complexity across the regions provides a comprehensive dataset to validate the effectiveness of the Volterra ROM in capturing the challenging unsteady aerodynamic responses. Blue and orange circles indicate the conditions plotted in Figure~\ref{fig_volterra_step_linear_nonlinear}.

The excitation signal for the step responses is a "smoothed step" of the pitch angle around an axis through the wing mid--chord:

\begin{equation}
    \alpha(\tau) = \alpha_0 \left( 1-e^{-\tau / \tau_{ref}}\right) 
\end{equation}
where $\tau$ is the non--dimensional time and $\tau_{ref}$ is equal to 0.6. This value is the result of a trial and error process aiming at reducing the impulsive response without affecting the circulatory part excessively. As a matter of fact the smoothing of the step input signal aims at cutting the higher part of the signal spectrum. This is not strictly necessary, but it helps achieve smoother CFD responses without affecting the frequency range used for the investigation.
We generated pitch step--responses of 1 $[deg]$ and 2 $[deg]$ in order to cover the linear and nonlinear dynamics.

In the region II of moderate $M-\alpha_0$ (see Figure~\ref{fig_sampled_points_1}) the response of the system tends to converge to the steady--state value without large oscillations (blue line in Figure~\ref{fig_volterra_step_linear_nonlinear}). On the other hand, when $M>0.80$ and $\alpha_0>3$ $[deg ]$ in the region III, the response is characterized by the appearance of a dynamic stall which leads to an initial increase in $C_{L}$ followed by a decrease toward the steady--state value (orange line in Figure~\ref{fig_volterra_step_linear_nonlinear}).

The conditions \(M = 0.829\) and \(\alpha_0 = 4.277\, [deg]\) and \(M = 0.745\) and \(\alpha_0 = 2.105\, [deg]\), plotted in Figure~\ref{fig_volterra_step_linear_nonlinear}, were selected to demonstrate the ROM ability to handle diverse aerodynamic scenarios. These points represent different regions in the $M-\alpha_0$ space—one near the upper boundary and the other closer to the center. The condition \(M = 0.829\) and \(\alpha_0 = 4.277\, [deg]\) was chosen for its complex flow phenomena, including shock waves and significant boundary layer separations, testing the ROM ability to capture highly nonlinear effects. Conversely, \(M = 0.745\) and \(\alpha_0 = 2.105\, [deg]\) is in a more moderate flow regime, where the effects of compressibility and flow separation are less pronounced, serving as a baseline for assessing the ROM performance in mildly unsteady conditions.

\subsection{Machine Learning for Volterra Kernels Identification} \label{subsec_ML_volterra_kernels}

\begin{figure} [!htb] 
    \centering
    \includegraphics[trim=1.5 1.5 1.5 1.5, clip, width=0.98\textwidth]{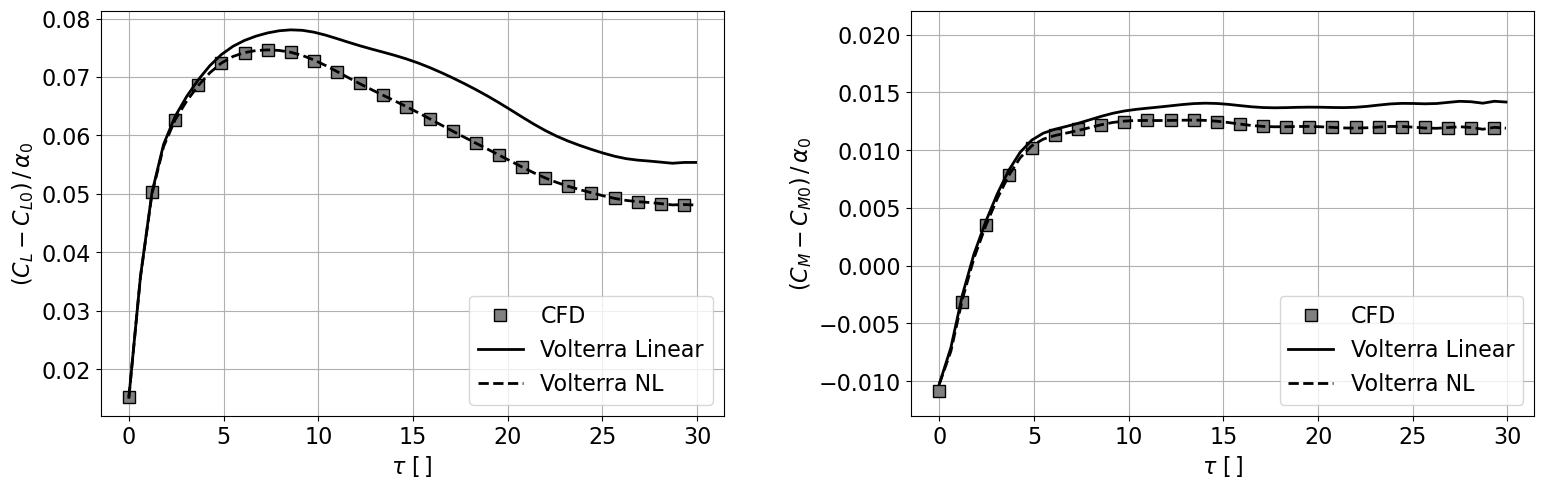}
    \caption{Aerodynamic loads reconstruction with linear and nonlinear Volterra kernels at $M = 0.829$ and $\alpha_0 = 4.277$ $[deg]$. The response was normalized by the pitch angle input step, after subtracting the steady--state value.} 
    \label{fig_nonlinear_contribute_CL}
\end{figure}

To evaluate the performance of the Volterra series, we first reconstructed $C_L$ and $C_M$ responses to a step input at \(M = 0.829\) and \(\alpha_0 = 4.277\) $[deg]$. As shown in Figure~\ref{fig_nonlinear_contribute_CL}, the nonlinear Volterra kernels closely match the CFD data, confirming the Volterra series capability to capture complex aerodynamic behavior.




\begin{figure}[!htb]
\centering
\subfigure[{Linear Volterra kernels for step--response of 1 [deg]} ]  {\label{linear_volterra_kernels_low_mach} \includegraphics[trim=0 0 0 2.8cm, clip, width=0.98\textwidth]{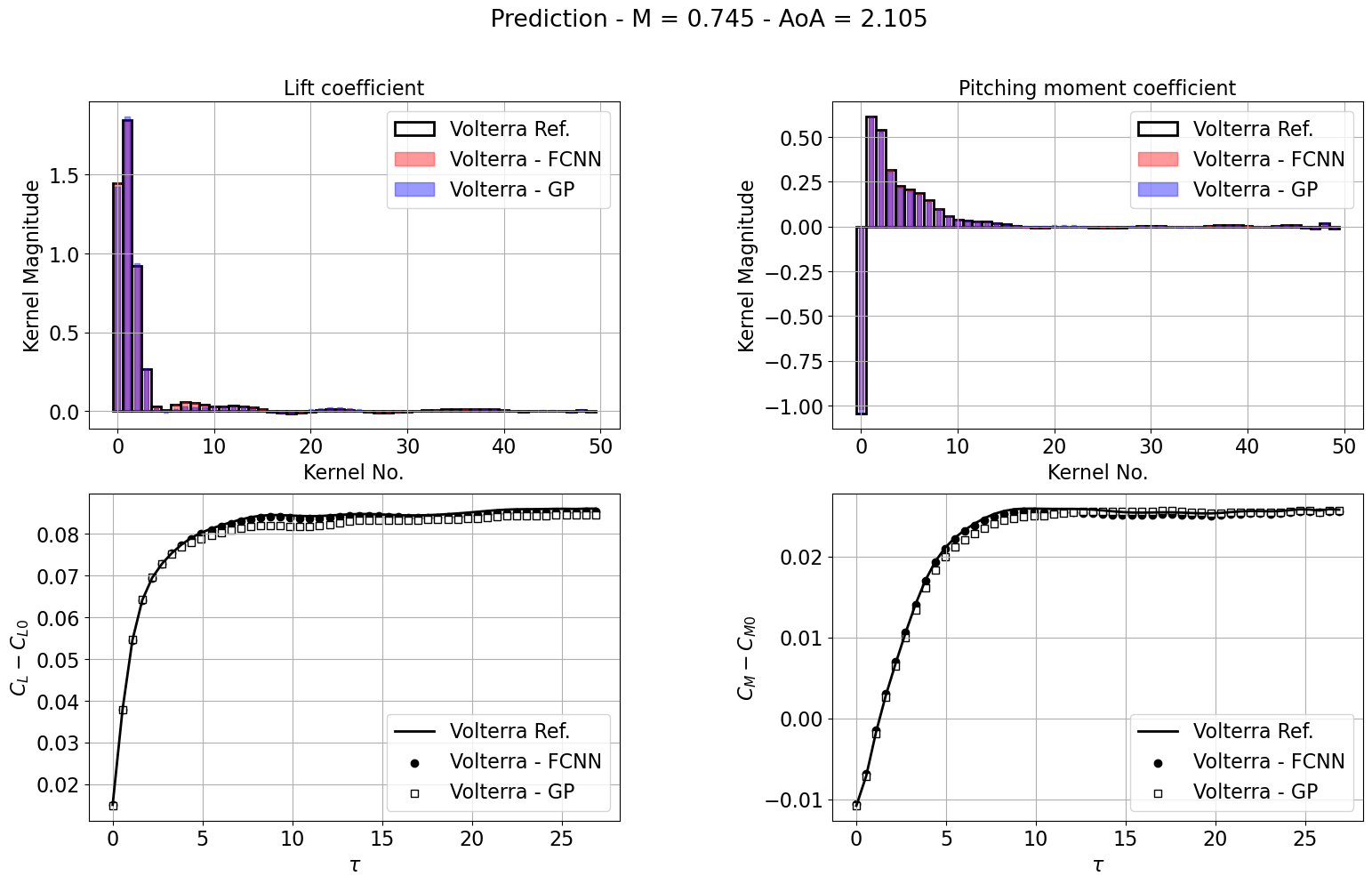}} \\
\subfigure[{NL Volterra kernels for step--response of 2 [deg]} ] {\label{NL_volterra_kernels_low_mach}
  \includegraphics[trim=0 0 0 2.8cm, clip, width=0.98\textwidth]{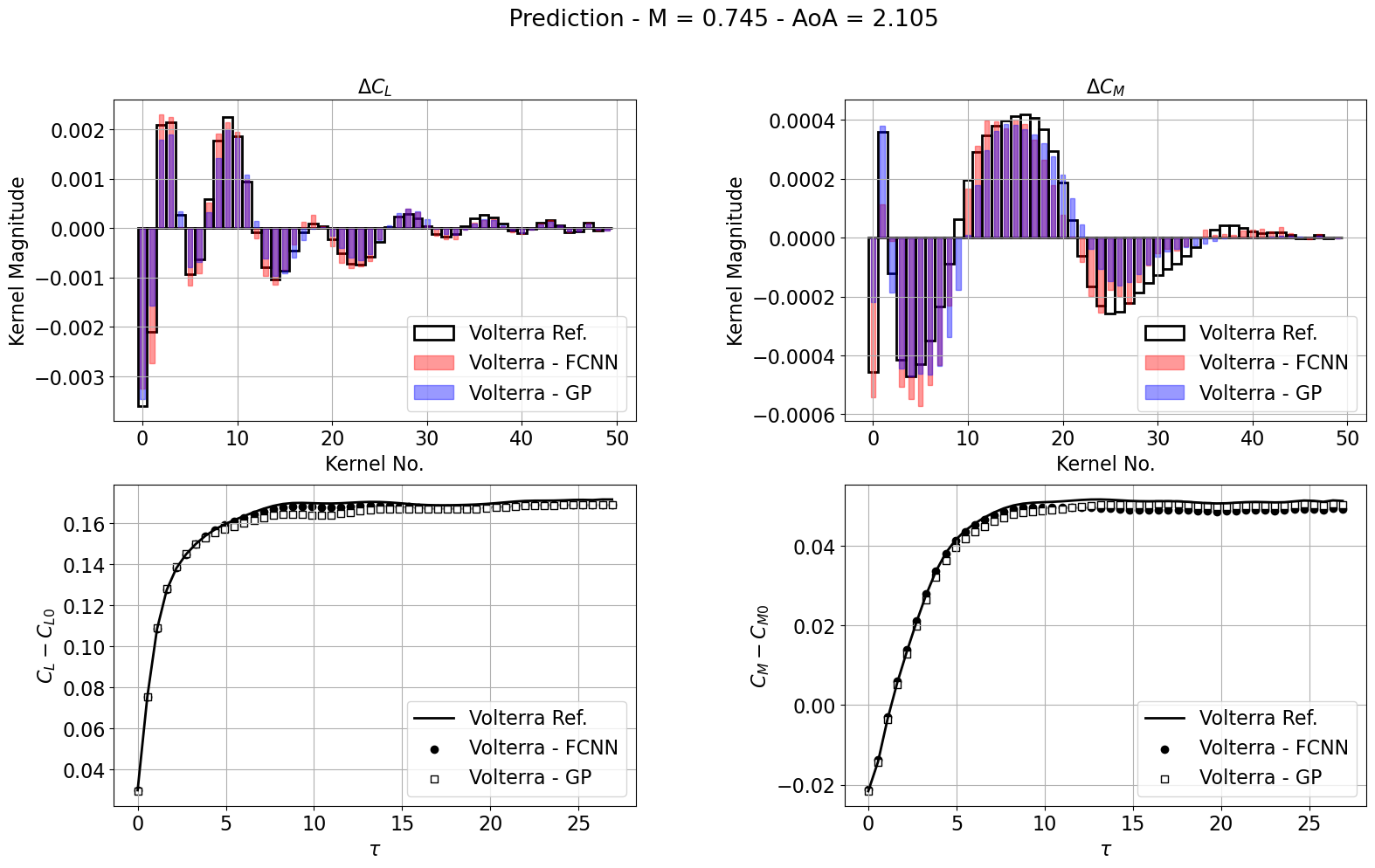}}
  
\caption{Volterra kernels predicted using ML methods and corresponding aerodynamic loads reconstructed at $M = 0.745$ and $\alpha_0 = 2.105$ $[deg]$. Continuous lines indicate the reference Volterra series derived from CFD data, while circles and squares denote the reconstructed responses using kernels predicted by FCNN and GPR, respectively.}
\label{fig_volterra_kernel_linear_NL_low_mach}
\end{figure}

\begin{figure}[!htb]
\centering
\subfigure[{Linear Volterra kernels for step--response of 1 [deg]} ] {\label{linear_volterra_kernels_high_Mach}  \includegraphics[trim=0 0 0 2.8cm, clip, width=0.98\textwidth]{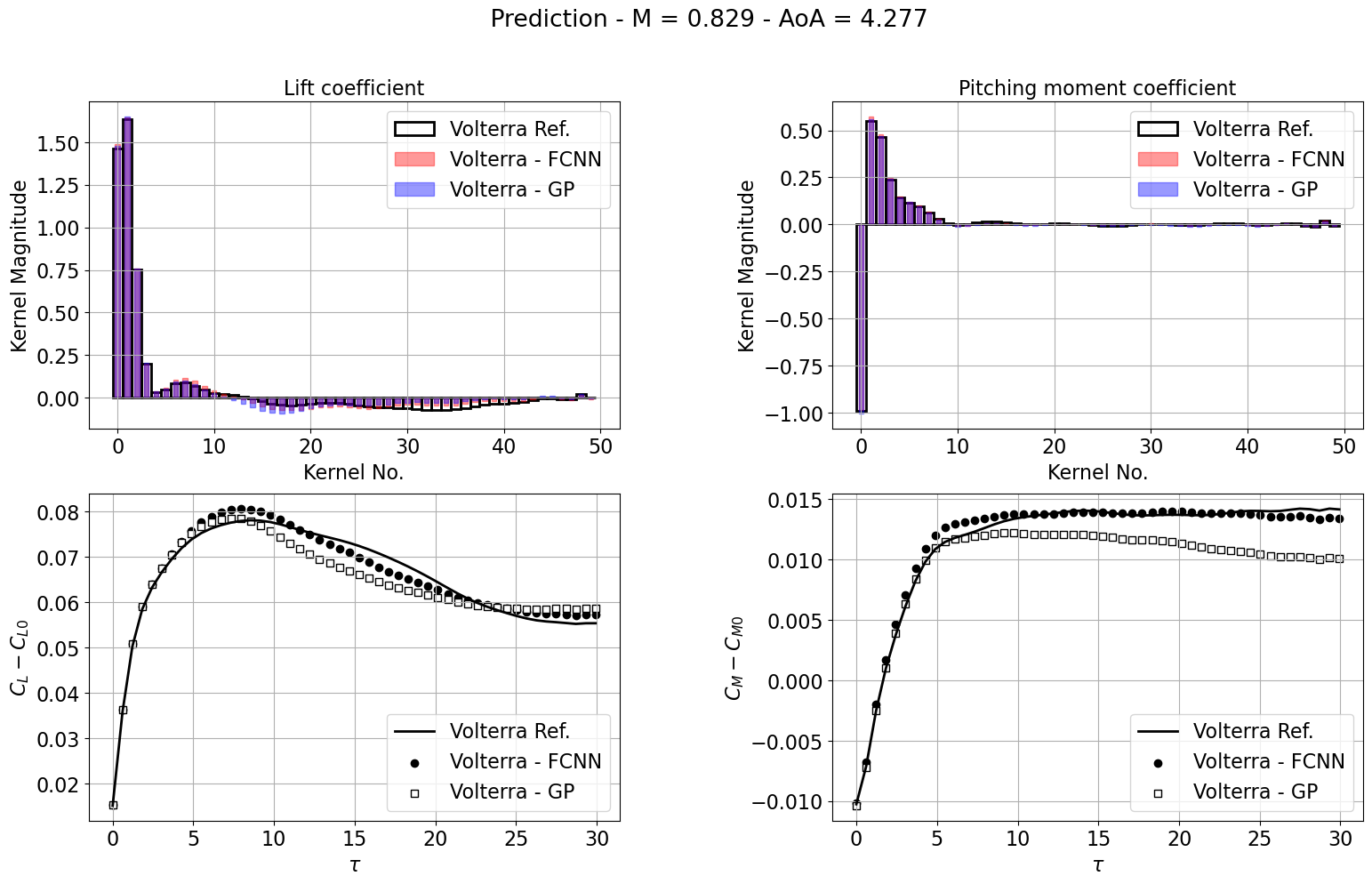}} \\
\subfigure[{NL Volterra kernels for step--response of 2 $[deg]$ }] {\label{NL_volterra_kernels_high_Mach}
  \includegraphics[trim=0 0 0 2.8cm, clip, width=0.98\textwidth]{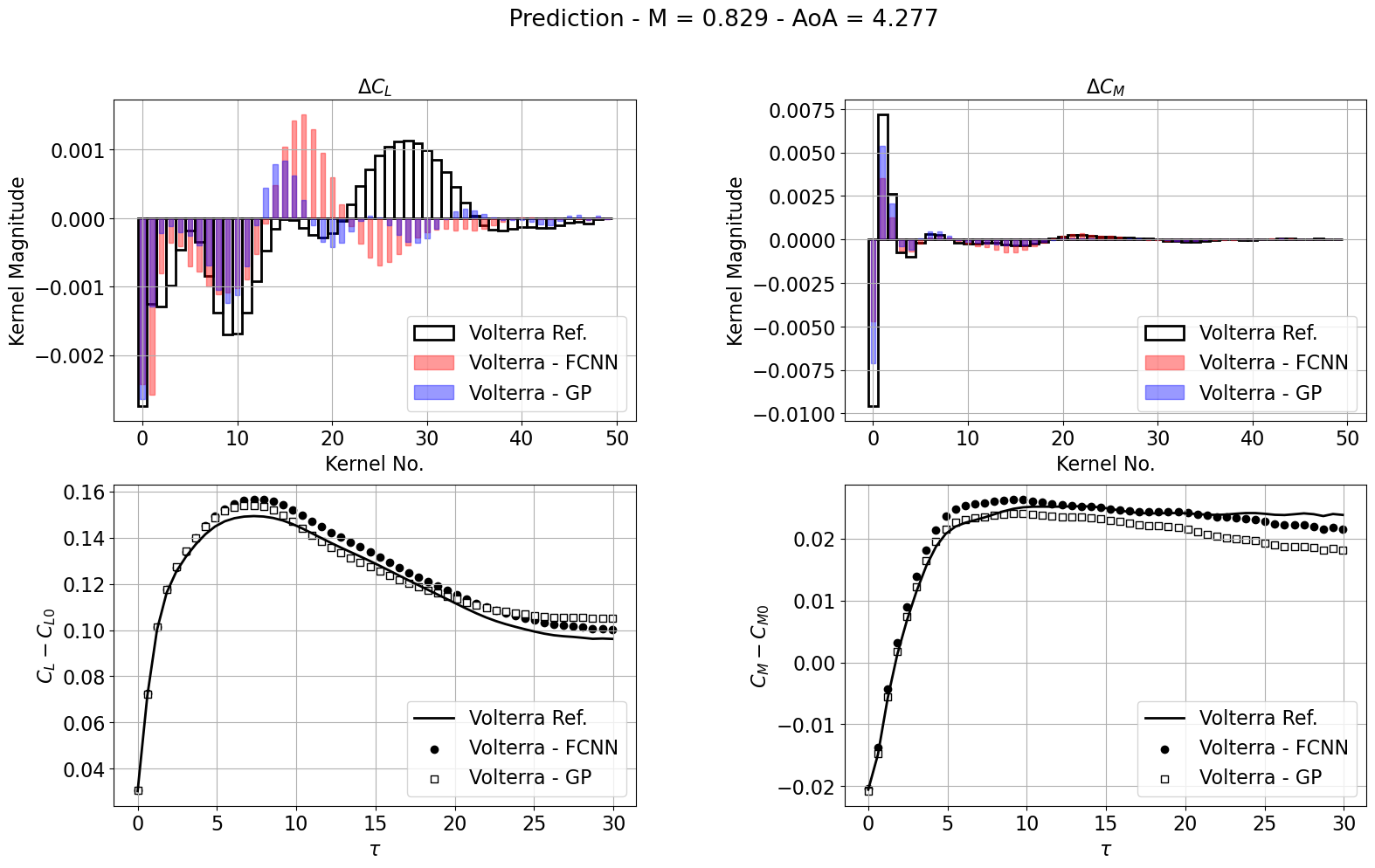}}
    \caption{Volterra kernels predicted using ML methods and corresponding aerodynamic loads reconstructed at $M = 0.829$ and $\alpha_0 = 4.277$ $[deg]$. Continuous lines indicate the reference Volterra series derived from CFD data, while circles and squares denote the reconstructed responses using kernels predicted by FCNN and GPR, respectively.}
    \label{fig_volterra_kernel_linear_NL_high_mach}
\end{figure}

Building on the methodology introduced in earlier sections, we employed two ML algorithms (FCNN and GPR) to reconstruct the Volterra kernels across the design space. The Volterra kernels reconstruction with the two ML models  at $M = 0.745$ and $\alpha_0 = 2.105$ $[deg ]$ and at $M = 0.829$ and $\alpha_0 = 4.277$ $[deg ]$ are reported in Figure~\ref{fig_volterra_kernel_linear_NL_low_mach} and Figure~\ref{fig_volterra_kernel_linear_NL_high_mach}, respectively. In Panel A, the aerodynamic loads at 1 $[deg]$ are reconstructed using linear Volterra kernels, while Panel B displays the reconstruction of aerodynamic loads at 2 $[deg]$ using nonlinear Volterra kernels. Both ML methods effectively capture the key characteristics of the kernels, showing good alignment with the reference Volterra kernels. When the responses are reconstructed using these kernels, they exhibit strong agreement with the CFD data, particularly at the moderate $M$ and \(\alpha_0\) (Figure~\ref{fig_volterra_kernel_linear_NL_low_mach}).
It is noteworthy that the linear kernels decrease in magnitude as the steady--state of the step--response is approached, whereas the nonlinear kernels have larger magnitudes also at higher kernel numbers as the nonlinear content is at low frequency. However, the situation changes at high $M$ and $\alpha_0$ (Figure~\ref{fig_volterra_kernel_linear_NL_high_mach}), where the FCNN outperforms the GPR in both linear and nonlinear cases. The network ability to regularize the data is a key feature that should be highlighted. The reconstructed aerodynamic loads appear smoother with fewer oscillations as the network filtered out the low frequency content. Based on these advantages, we have chosen to reconstruct Volterra kernels with FCNN for the two proposed applications.

\subsection{Identification of Pitch Step--responses} \label{subsec_identification_pitch_step}

\begin{figure} [!htb] 
    \centering
    \includegraphics[trim=0 0 0 0, clip, width=0.58\textwidth]{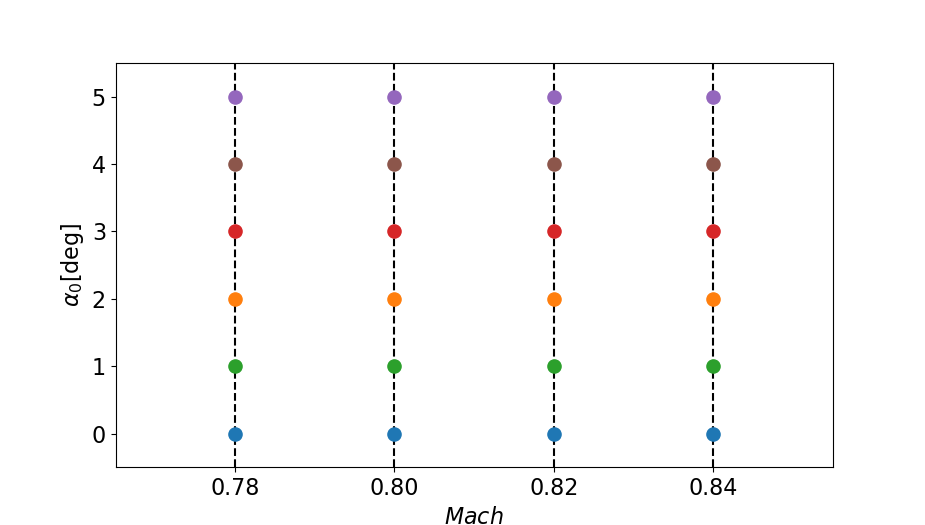}
    \caption{Parametric space of $M$ and $\alpha_0$ used to identify pitch step-responses with the Volterra ROM.} 
    \label{fig_mach_vs_aoa_step}
\end{figure}

We exploit the Volterra ROM methodology developed and the training data presented in the third test case (section \ref{subsec_ML_volterra_kernels}) to reconstruct the pitch step response in a series of $\alpha_0$ sweeps at different $M$ (Figure~\ref{fig_mach_vs_aoa_step}).

\begin{figure}[!htb]
\begin{center}
\subfigure[{$M = 0.78$} ]  {\includegraphics[trim=0 0 1.4cm  1cm, clip, width=0.49\textwidth]{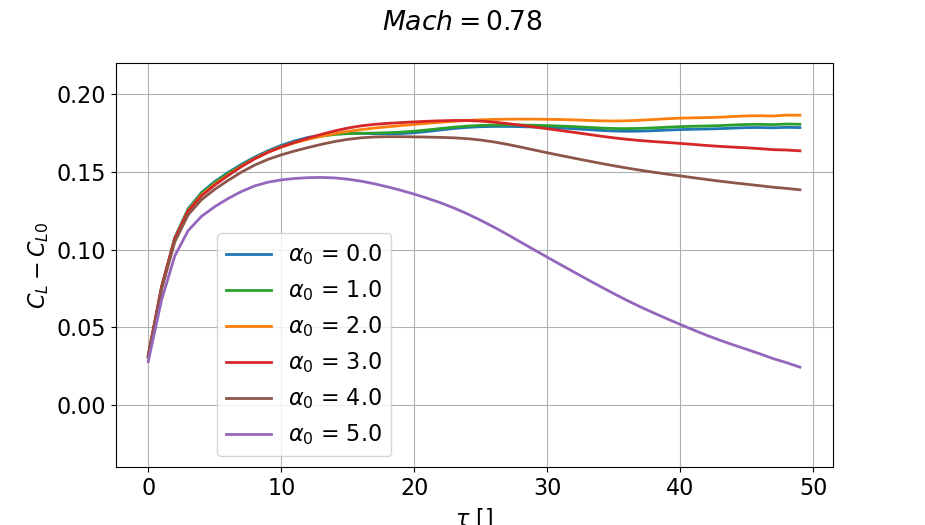}} 
\subfigure[{$M = 0.80$} ] { \includegraphics[trim=0 0 1.4cm  1cm, clip, width=0.49\textwidth]{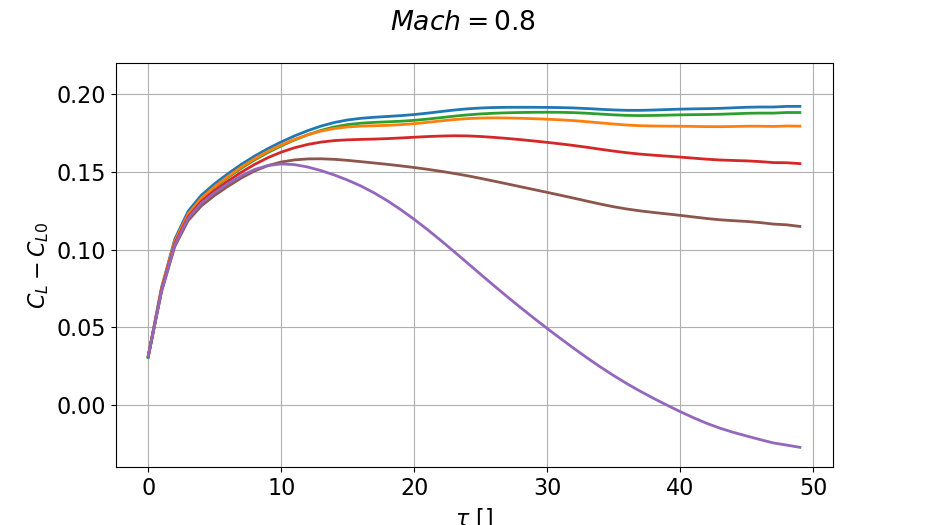}} \\
  \subfigure[{$M = 0.82$} ]  {\includegraphics[trim=0 0 1.4cm  1cm, clip, width=0.49\textwidth]{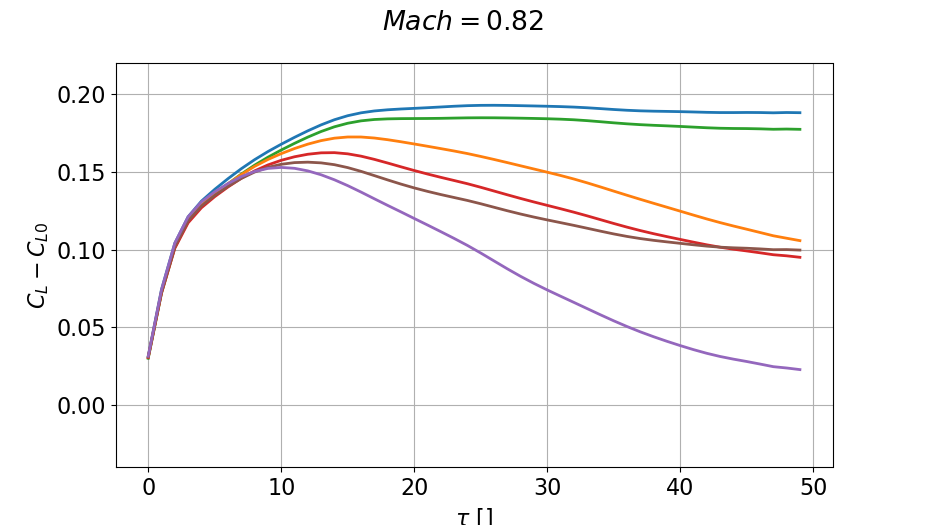}} 
\subfigure[{$M = 0.84$} ] {\includegraphics[trim=0 0 1.4cm  1cm, clip, width=0.49\textwidth]{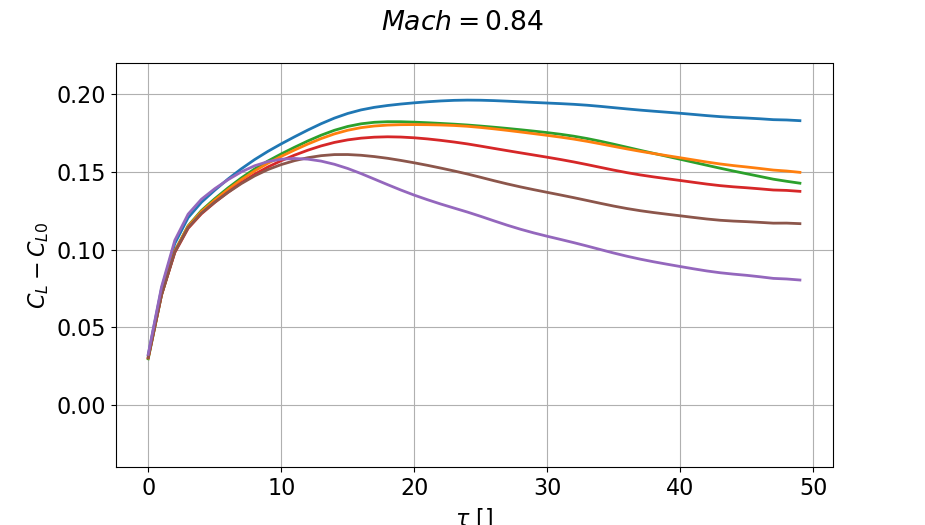}}
\caption{Pitch step--responses of 2 $[deg]$ for several $\alpha_0$ at four distinct fixed $M$ reconstructed using nonlinear Volterra kernels predicted by FCNN model.}
\label{fig_Pitch_step_responses_mach_fix}
\end{center}
\end{figure}

The identified step--responses to a pitch input angle of 2 $[deg]$ are reported in Figure~\ref{fig_Pitch_step_responses_mach_fix}. Each panel consists of six distinct responses constructed maintaining fixed $M$ and varying $\alpha_0$ $\in [0,5]$ $[deg]$. As we gradually increase $\alpha_0$, the step-responses demonstrate a distinct dynamic stall behavior. Notably, this behavior emerges at progressively lower $\alpha_0$ as $M$ increases, indicating a strong influence of $M$ on the onset of dynamic stall. This is particularly evident at higher $M$ values, where flow separation becomes more extensive, and the interaction between shock waves and the boundary layer becomes more significant. 
This analysis demonstrates the effectiveness of the methodology in capturing the patterns of the aerodynamic response. The identified step-responses provide valuable insights into how changes in $M$ and $\alpha_0$ influence the dynamic stall and flow separation. These results showcase the method capability to explore aerodynamic responses under a range of operating conditions, reinforcing its potential for practical applications in aerodynamic analysis.

\subsection{Reconstruction of Harmonic Signals} \label{subsec_harmonic_signal_reconstruction}

\begin{figure} [!htb] 
    \centering
        \includegraphics[trim= 0 0 0 0, clip, width=0.49\textwidth]{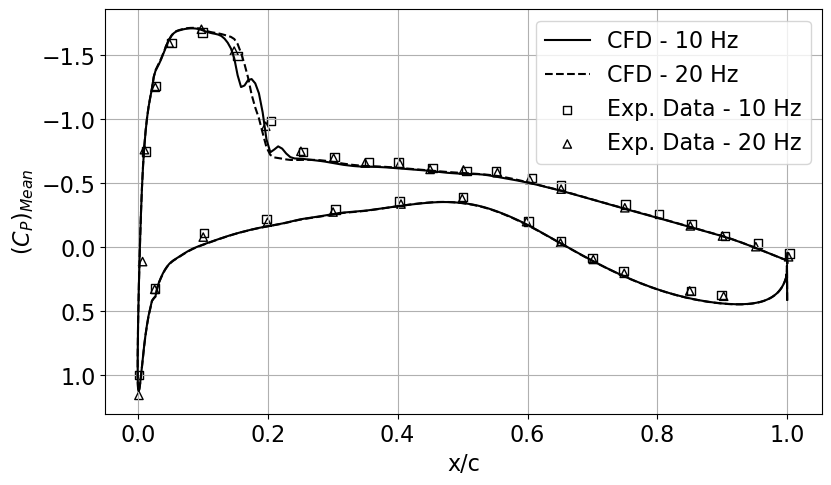} \\
    \includegraphics[trim= 0 0 0 0, clip, width=0.98\textwidth]{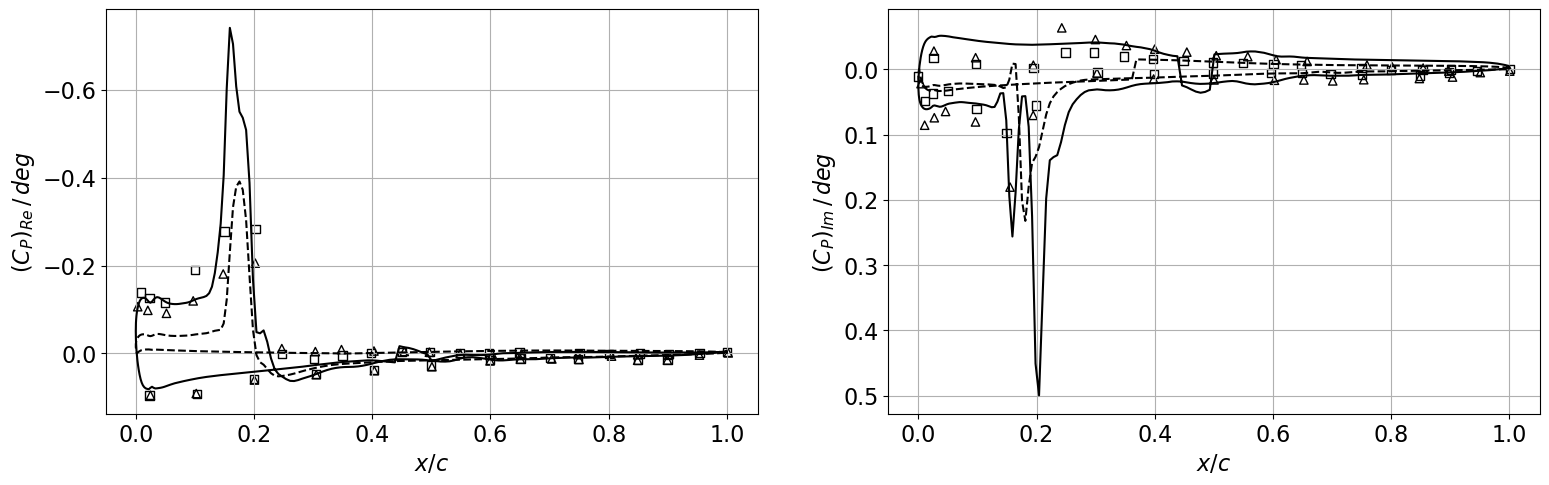}
    \caption{Mean, real and imaginary parts of the pressure coefficient at peak frequency at $60\%$ of the span ($\bar{\alpha}= 5$ [deg], $M= 0.70$ and $q = 170$ [psf]).}
    \label{fig_CP_validation_harmonic_signal}
\end{figure}

As a second application of the Volterra ROM methodology and the training data presented in section~\ref{subsec_ML_volterra_kernels}, we predict the response to a harmonic signal at  $M= 0.70$, $\bar{\alpha}= 5$ $[deg ]$ and $q = 170 \, psf$. We generated two distinct responses, one with a frequency of $10$ Hz and an amplitude of $\alpha_A=1.03$ deg, and another with a frequency of $20$ Hz and an amplitude of $\alpha_A=0.50$ deg. The validation conditions were based on Piatak et al. experimental data~\cite{piatak2003oscillating}. We initially validated our CFD data by computing the real and imaginary components of $C_P/deg$ by performing the Fast Fourier Transform (FFT) and taking the peak frequency, and comparing them to the findings of Piatak et al.~\cite{piatak2003oscillating}. As shown in Figure~\ref{fig_CP_validation_harmonic_signal}, we observed a good agreement between our CFD and experimental data. It is interesting to note that the pitch oscillation induces a forward and backward motion of the shock wave, resulting in significant variations in both the real and imaginary components within that specific region. These observations are in line  with the findings reported by Heeg et al.~\cite{heeg2016overview}.

\begin{figure}[!htb]
\begin{center}
\subfigure[$f=10 \, Hz - \alpha_A=1.03$ deg] {\label{harmonic_signal_10hz}  \includegraphics[trim=0cm 0cm 1.4cm 1cm, clip, width=0.49\textwidth]{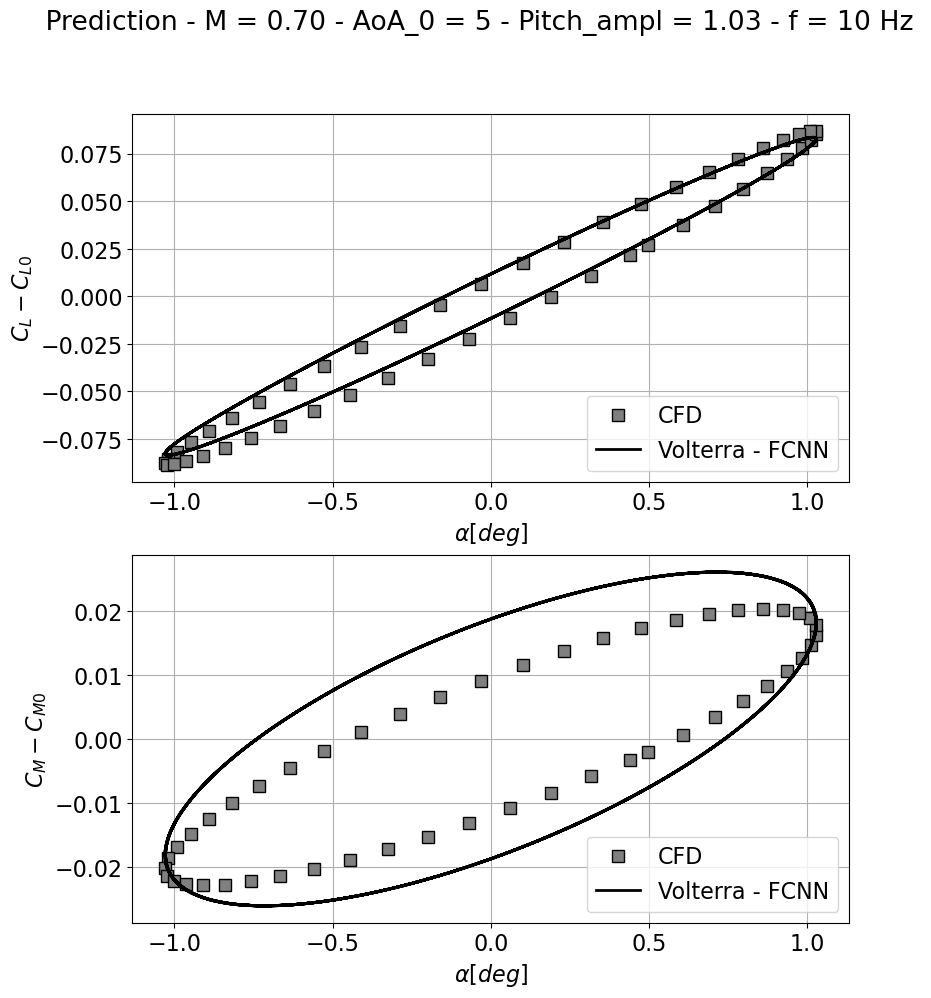}} 
\subfigure[$f=20 \, Hz - \alpha_A=0.50$ deg] {\label{harmonic_signal_20hz}
  \includegraphics[trim=0cm 0cm 1.4cm 1cm, clip, width=0.49\textwidth]{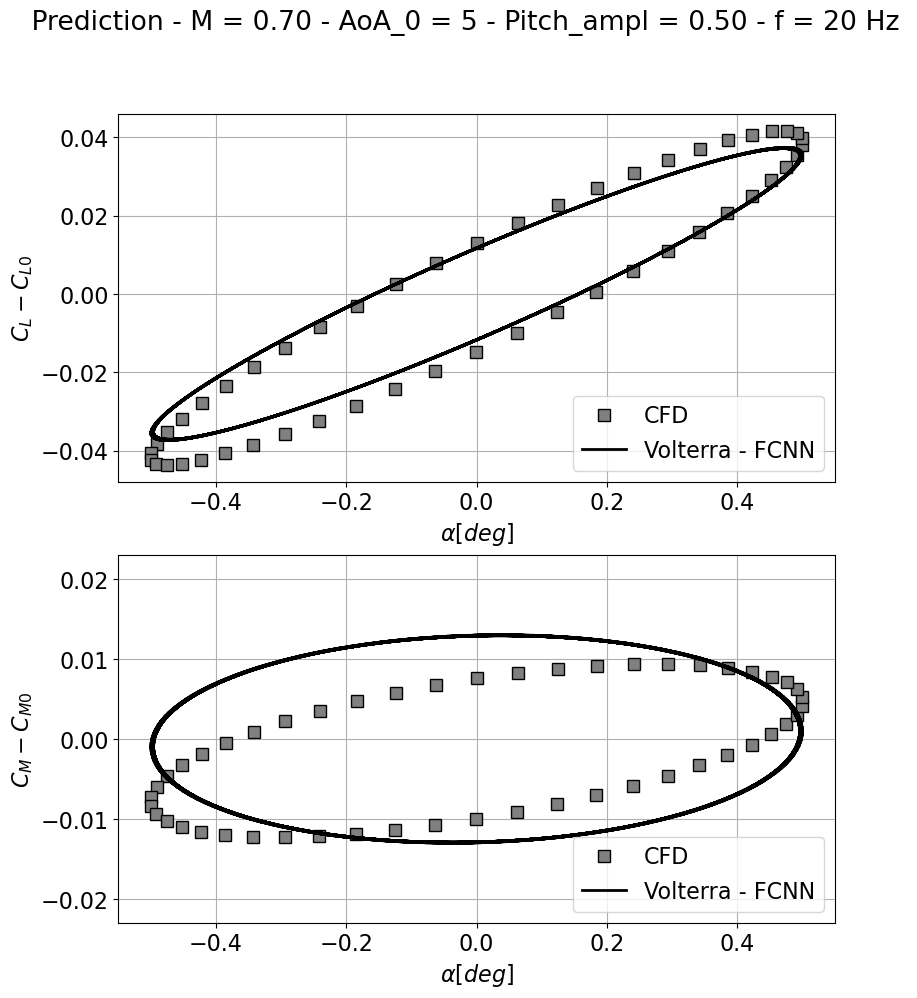}}
    \caption{Harmonic signal reconstruction with Volterra ROM at $\bar{\alpha}= 5$ [deg], $M= 0.70$ and $q = 170$ [psf]. CFD data are represented by square markers, while continuous line denotes the reconstructed responses using nonlinear Volterra kernels predicted by FCNN.}
    \label{fig_harmonic_signal_reconstruction}
    \end{center}
\end{figure}

After validating the CFD data, the reconstruction of $C_L$ and $C_M$ can be performed using the Volterra ROM as illustrated in Figure~\ref{fig_harmonic_signal_reconstruction}. The Volterra kernels are identified using FCNN and then convolved with the input signal to obtain the predicted results. The reconstruction of the harmonic signal is demonstrated at two frequencies, 10 Hz and 20 Hz. However, when predicting at high $\alpha_0$, the accuracy of the results can be challenging due to the global instability of the flowfield, which is associated with boundary layer separation. Nevertheless, the Volterra ROM closely follows the curves.

\begin{figure} [!htb] 
    \centering
    \includegraphics[trim= 0cm 0cm 0cm 1cm, clip, width=0.98\textwidth]{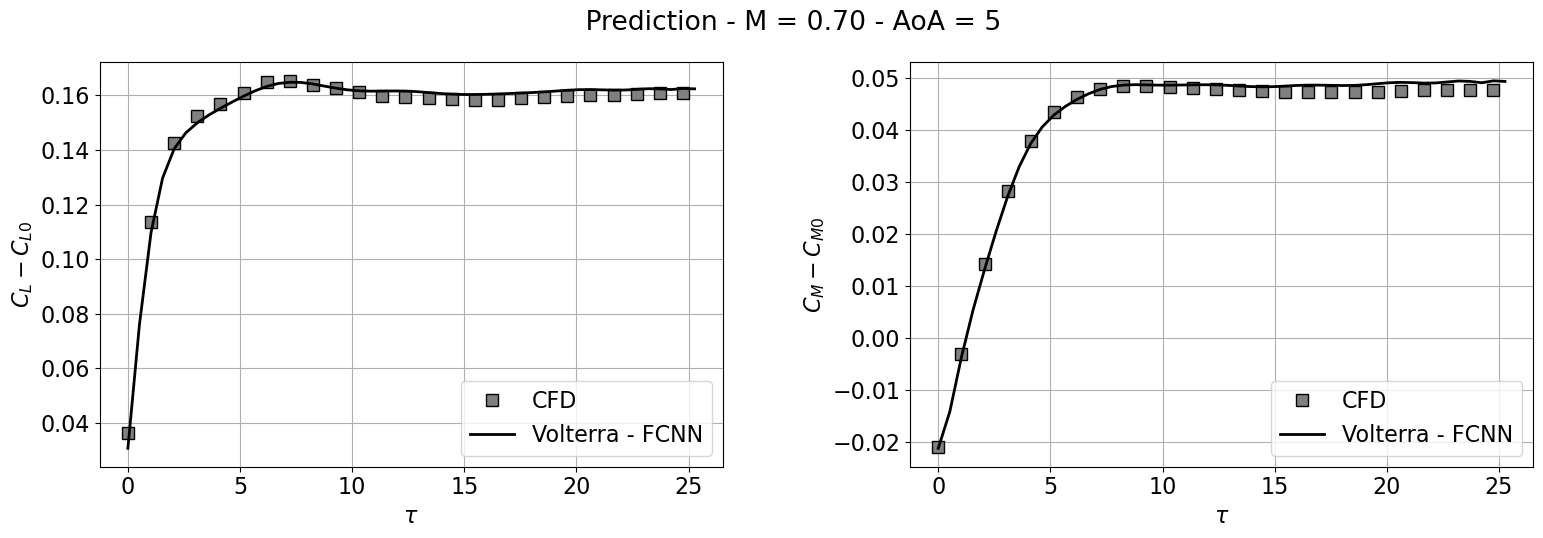}
    \caption{Step--response prediction at $\alpha_0 = 5$ [deg], $M= 0.70$ and $q = 170$ [psf]. CFD data are represented by square markers, while continuous line denotes the reconstructed responses using nonlinear Volterra kernels predicted by FCNN.}
    \label{fig_step_M070_AoA5}
\end{figure}

The method presented involves an "offline" identification of Volterra kernels using step--responses. These kernels are then applied to different frequency values of harmonic inputs to achieve dependable estimates of nonlinear loads. The reliability of these estimations can be evaluated by comparing the identified model to the step--response data. The robustness of this approach ensures accurate nonlinear estimation, as long as the step--responses match the desired levels closely (refer to Figure~\ref{fig_step_M070_AoA5}).

\begin{figure}[!htb]
\begin{center}
\subfigure[{$M = 0.78$} ]  {\includegraphics[trim=0 0 1.4cm 1cm, clip, width=0.49\textwidth]{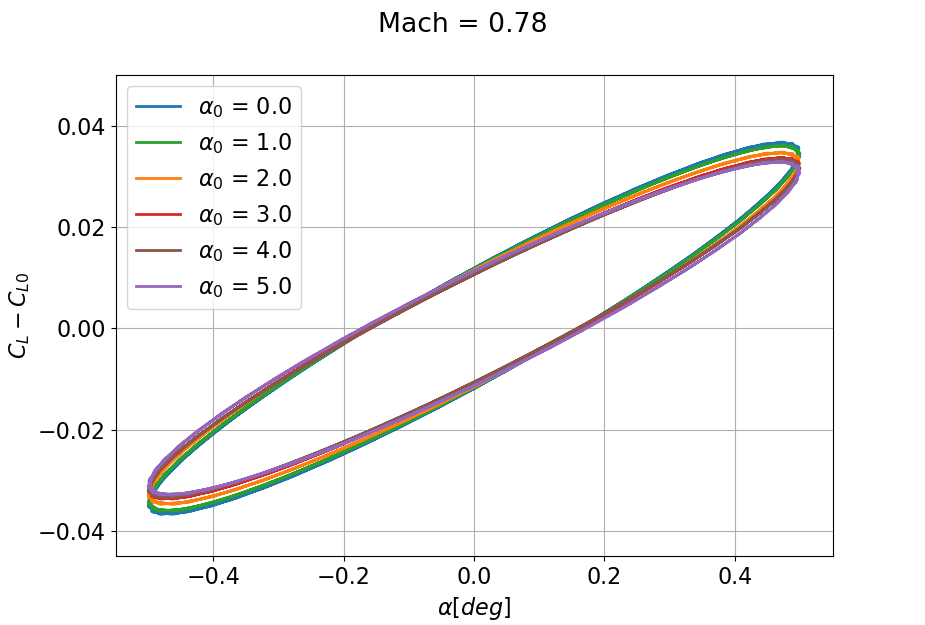}} 
\subfigure[{$M = 0.80$} ] { \includegraphics[trim=0 0 1.4cm 1cm, clip, width=0.49\textwidth]{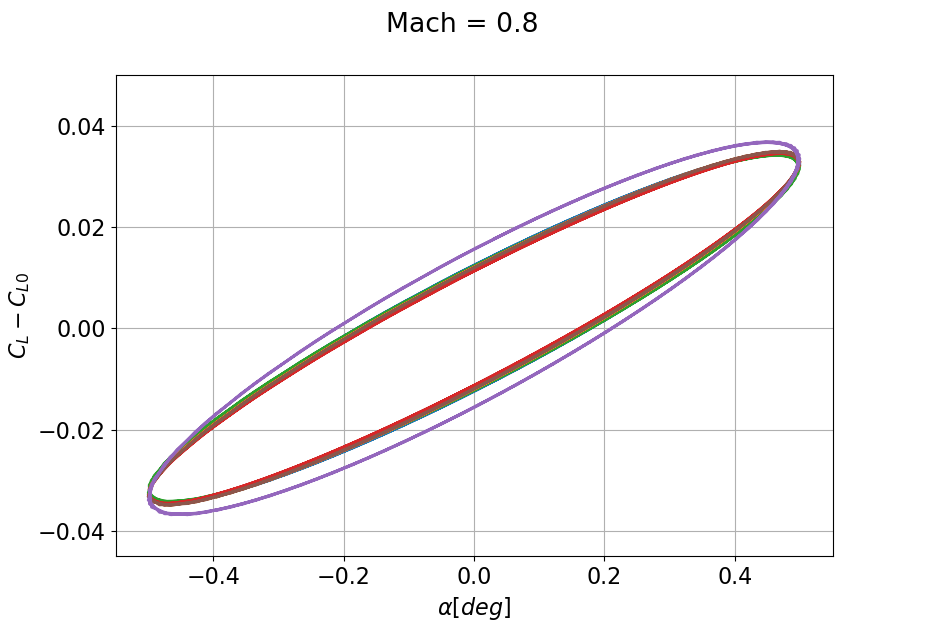}} \\
  \subfigure[{$M = 0.82$} ]  {\includegraphics[trim=0 0 1.4cm 1cm, clip, width=0.49\textwidth]{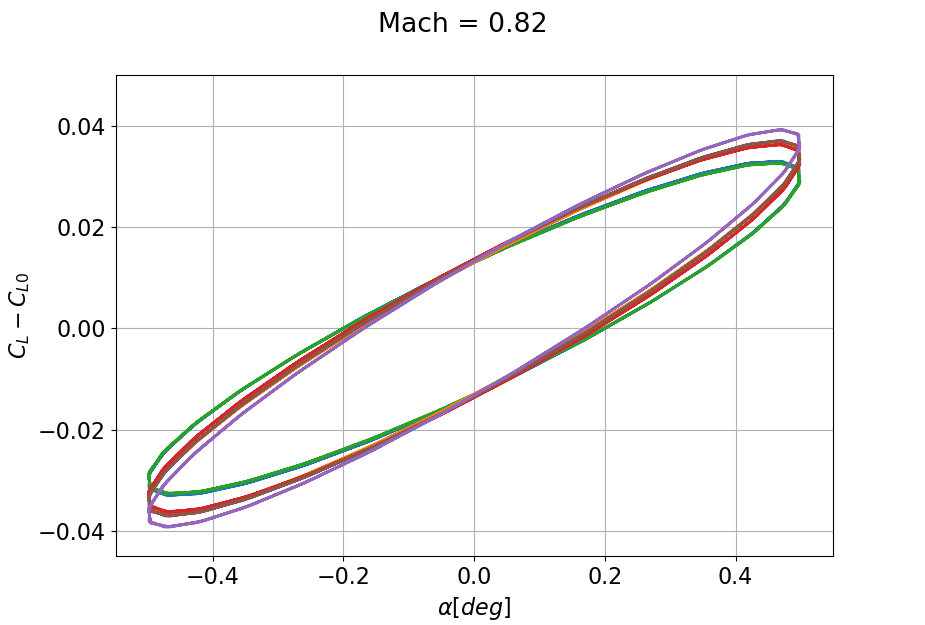}} 
\subfigure[{$M = 0.84$} ] {\includegraphics[trim=0 0 1.4cm 1cm, clip, width=0.49\textwidth]{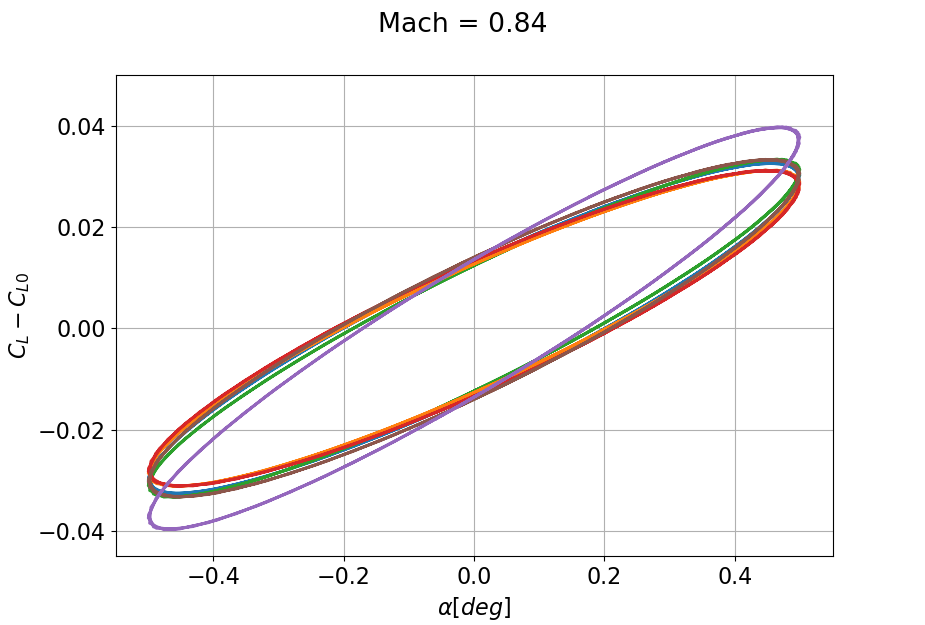}}
\caption{Harmonic signal reconstruction for different $\alpha_0$ at fixed $M$ ($f=20$ [Hz] - $\alpha_A=0.50$ [deg]) using nonlinear Volterra kernels predicted by FCNN model.}
\label{fig_harmonic_signals_mach_fix}
\end{center}
\end{figure}

Finally, we present four $M$ sweeps in Figure~\ref{fig_mach_vs_aoa_step}. As $\alpha_0$ increases, we observe a corresponding increase in the magnitude of hysteresis, as depicted in Figure~\ref{fig_harmonic_signals_mach_fix}. These findings are consistent with the experimental results reported by Heeg et al.~\cite{heeg2013experimental}, further validating our observations.

\subsection{Computing Cost Analysis}

A detailed computational cost analysis was performed to evaluate the efficiency of the Volterra ROM in comparison to the high-fidelity CFD simulations. Table~\ref{tab_cpu_time_comparison} summarizes the costs for training and prediction. In the 2D case, CFD requires 126,000 CPU hours for 140 runs, with a single run taking 900 hours. In contrast, the Volterra ROM with the GPR model required 0.01 GPU hours for training and only 0.0003 GPU hours (approximately 1 second) for a single prediction. The FCNN model in the same 2D case involved 1.6 GPU hours for optimization, 0.03 hours for training, and similarly 0.0003 GPU hours for a single prediction. In a practical application, the virtually instantaneous predictions could provide significant computational savings. The 3D case demonstrates an even larger discrepancy, as CFD requires 560,000 CPU hours for 140 runs for training (4,000 hours for a single run). The Volterra ROM with the GPR model needed just 0.01 GPU hours for training and 0.0003 GPU hours for prediction, while the FCNN model required 1.6 GPU hours for optimization, 0.03 GPU hours for training, and the same 0.0003 GPU hours for prediction.

This comparison highlights the efficiency and scalability of the Volterra ROM, providing accurate predictions with drastically lower computational demands compared to traditional CFD simulations.

Each CFD simulation was performed on a high-performance computing system with an Intel Skylake-based architecture, utilizing 3 nodes with 40 CPU cores each. In contrast, the training of the Volterra model was executed on an Intel XEON W-2255 CPU paired with a NVIDIA RTX A4000 GPU.

\begin{table}[!htb]
\small
\centering
\begin{tabular}{l c c c c c c c}
\hline 
\hline 
\textbf{Test case} &  \multicolumn{2}{c}{\textbf{CFD (CPU hours)}} &  \multicolumn{4}{c}{\textbf{Volterra ROM (GPU hours)}} \\
\cmidrule(lr){2-3}\cmidrule(lr){4-7}
  & \multicolumn{2}{c}{Simulation}   & Model & Optimization & Training & Prediction \\
  & (140 runs) & (1 run)  & Type & (50 trials) & (1 model) &  (1 sample) \\
\hline
2D & 126,000 &  900 &  GPR & -- &  0.01 &  0.0003 ($\sim{1s}$) \\
   &         &      &  FCNN & 1.6 &  0.03 &  0.0003 ($\sim{1s}$) \\
\hline
3D & 560,000 &  4,000 &  GPR & -- &  0.01 &  0.0003 ($\sim{1s}$) \\
   &         &        &  FCNN & 1.6 &  0.03 &  0.0003 ($\sim{1s}$) \\
\hline
\hline 
\end{tabular}
\caption{Computing cost comparison between Volterra ROM and CFD for the two test cases. 140 runs include 70 for small-amplitude input and 70 for large-amplitude input.}
\label{tab_cpu_time_comparison}
\end{table}

\section{Conclusions} \label{sec_conclusions}

Predicting steady and unsteady aerodynamic characteristics in the transonic regime is needed for the development of high-speed aircraft. A compromise between lowering the computing costs and increasing the prediction accuracy is generally found in a ROM representation of the aerodynamics. Many are the possible representations, but none has become the standard. We considered a ROM based on Volterra series for a single-input, single-output system.

The series contains first-order (linear) and second-order (nonlinear) Volterra kernels to capture adequately the physical complexity resulting from unsteady, transonic flows around two and three-dimensional configurations. A two-steps process is used to identify the Volterra kernels, justified in light of the dominant character of the linear component in any nonlinear response. First, the linear kernels are identified from a response obtained with a small-amplitude input. Then, the nonlinear kernels are identified from the difference between a large-amplitude input response and the small-amplitude input response. To increase the readiness of the approach towards future applications, for example aeroelasticity and optimisation, we considered the development of a parametric Volterra series by deploying two ML algorithms across the $M - \alpha_0$ design space.

This study demonstrates the effectiveness of the Volterra series in accurately capturing both linear and nonlinear aerodynamic responses in the transonic regime. Using GPR and, more successfully, FCNN to reconstruct linear and nonlinear Volterra kernels, the methodology reproduced complex phenomena such as dynamic stall across a wide range of $M$ and $\alpha_0$. The ML-based interpolation allowed the model to generalize effectively within the parametric space, handling diverse aerodynamic conditions with high accuracy. The analysis of the dynamic response via Volterra ROM may allow substantial savings in computational cost compared to CFD, as the training only exploited two responses per flow condition and the predictions can be produced in the entire parameters space at virtually no computational costs. The approach proved scalable for both two and three-dimensional configurations, confirming its potential for more complex aerodynamic systems.

Future work shall be directed towards extending the methodology to multi-input systems. While straightforward in its implementation, the identification of the nonlinear kernels require system responses where the inputs are simultaneously applied to capture any interaction.

\section*{Acknowledgement}

The authors acknowledge the use of the IRIDIS High Performance Computing Facility, and associated support services at the University of Southampton, in the completion of this work. The project has been supported by Digitalization Initiative of the Zurich Higher Education Institutions (DIZH) grant from Zurich University of Applied Sciences (ZHAW).

\bibliography{sample}

\begin{thebibliography}{42}
\newcommand{\enquote}[1]{``#1''}
\providecommand{\natexlab}[1]{#1}
\providecommand{\url}[1]{\texttt{#1}}
\providecommand{\urlprefix}{URL }
\expandafter\ifx\csname urlstyle\endcsname\relax
  \providecommand{\doi}[1]{\discretionary{}{}{}https://doi.org/#1}\else
  \providecommand{\doi}[1]{\discretionary{}{}{}\urlstyle{rm}\url{https://doi.org/#1}}\fi

\bibitem[{Dowell(2023)}]{dowell2023reduced}
Dowell, E., \enquote{Reduced-order modeling: a personal journey,} \emph{Nonlinear Dynamics}, Vol. 111, No.~11, 2023, pp. 9699--9720.

\bibitem[{Silva(1991)}]{silva1991methodology}
Silva, W., \enquote{A methodology for using nonlinear aerodynamics in aeroservoelastic analysis and design,} \emph{32nd Structures, Structural Dynamics, and Materials Conference}, 1991, p. 1110.

\bibitem[{Chatterjee(2000)}]{chatterjee2000introduction}
Chatterjee, A., \enquote{An introduction to the proper orthogonal decomposition,} \emph{Current science}, 2000, pp. 808--817.

\bibitem[{Schmid(2010)}]{schmid2010dynamic}
Schmid, P.~J., \enquote{Dynamic mode decomposition of numerical and experimental data,} \emph{Journal of fluid mechanics}, Vol. 656, 2010, pp. 5--28.

\bibitem[{Lassila et~al.(2014)Lassila, Manzoni, Quarteroni, and Rozza}]{lassila2014model}
Lassila, T., Manzoni, A., Quarteroni, A., and Rozza, G., \enquote{Model order reduction in fluid dynamics: challenges and perspectives,} \emph{Reduced Order Methods for modeling and computational reduction}, 2014, pp. 235--273.

\bibitem[{Lucia et~al.(2004)Lucia, Beran, and Silva}]{lucia2004reduced}
Lucia, D.~J., Beran, P.~S., and Silva, W.~A., \enquote{Reduced-order modeling: new approaches for computational physics,} \emph{Progress in aerospace sciences}, Vol.~40, No. 1-2, 2004, pp. 51--117.

\bibitem[{Amsallem and Farhat(2008)}]{amsallem2008interpolation}
Amsallem, D., and Farhat, C., \enquote{Interpolation method for adapting reduced-order models and application to aeroelasticity,} \emph{AIAA journal}, Vol.~46, No.~7, 2008, pp. 1803--1813.

\bibitem[{Amsallem et~al.(2009)Amsallem, Cortial, Carlberg, and Farhat}]{amsallem2009method}
Amsallem, D., Cortial, J., Carlberg, K., and Farhat, C., \enquote{A method for interpolating on manifolds structural dynamics reduced-order models,} \emph{International journal for numerical methods in engineering}, Vol.~80, No.~9, 2009, pp. 1241--1258.

\bibitem[{Amsallem et~al.(2012)Amsallem, Zahr, and Farhat}]{amsallem2012nonlinear}
Amsallem, D., Zahr, M.~J., and Farhat, C., \enquote{Nonlinear model order reduction based on local reduced-order bases,} \emph{International Journal for Numerical Methods in Engineering}, Vol.~92, No.~10, 2012, pp. 891--916.

\bibitem[{Zahr and Farhat(2015)}]{zahr2015progressive}
Zahr, M.~J., and Farhat, C., \enquote{Progressive construction of a parametric reduced-order model for PDE-constrained optimization,} \emph{International Journal for Numerical Methods in Engineering}, Vol. 102, No.~5, 2015, pp. 1111--1135.

\bibitem[{Amsallem and Farhat(2011)}]{amsallem2011online}
Amsallem, D., and Farhat, C., \enquote{An online method for interpolating linear parametric reduced-order models,} \emph{SIAM Journal on Scientific Computing}, Vol.~33, No.~5, 2011, pp. 2169--2198.

\bibitem[{Shaw and Pierre(1999)}]{shaw1999modal}
Shaw, S.~W., and Pierre, C., \enquote{Modal Analysis-Based Reduced-Order Models for Nonlinear Structures\&mdash; An Invariant Manifold Approach,} \emph{The shock and vibration digest}, Vol.~31, No.~1, 1999, pp. 3--16.

\bibitem[{Hall et~al.(2002)Hall, Thomas, and Clark}]{hall2002computation}
Hall, K.~C., Thomas, J.~P., and Clark, W.~S., \enquote{Computation of unsteady nonlinear flows in cascades using a harmonic balance technique,} \emph{AIAA journal}, Vol.~40, No.~5, 2002, pp. 879--886.

\bibitem[{Schuster et~al.(2003)Schuster, Liu, and Huttsell}]{schuster2003computational}
Schuster, D.~M., Liu, D.~D., and Huttsell, L.~J., \enquote{Computational aeroelasticity: success, progress, challenge,} \emph{Journal of Aircraft}, Vol.~40, No.~5, 2003, pp. 843--856.

\bibitem[{Silva(1999)}]{silva1999reduced}
Silva, W., \enquote{Reduced-order models based on linear and nonlinear aerodynamic impulse responses,} \emph{40th Structures, Structural Dynamics, and Materials Conference and Exhibit}, 1999, p. 1262.

\bibitem[{Volterra(1887)}]{volterra1887sopra}
Volterra, V., \emph{Sopra le funzioni che dipendono da altre funzioni}, Tipografia della R. Accademia dei Lincei, 1887.

\bibitem[{Dodd and Harrison(2002)}]{dodd2002new}
Dodd, T.~J., and Harrison, R.~F., \enquote{A new solution to Volterra series estimation,} \emph{IFAC Proceedings Volumes}, Vol.~35, No.~1, 2002, pp. 67--72.

\bibitem[{Koh and Powers(1985)}]{koh1985second}
Koh, T., and Powers, E., \enquote{Second-order Volterra filtering and its application to nonlinear system identification,} \emph{IEEE Transactions on acoustics, speech, and signal processing}, Vol.~33, No.~6, 1985, pp. 1445--1455.

\bibitem[{Marzocca et~al.(2004)Marzocca, Silva, and Librescu}]{marzocca2004nonlinear}
Marzocca, P., Silva, W.~A., and Librescu, L., \enquote{Nonlinear open-/closed-loop aeroelastic analysis of airfoils via Volterra series,} \emph{AIAA journal}, Vol.~42, No.~4, 2004, pp. 673--686.

\bibitem[{Loghmanian et~al.(2011)Loghmanian, Yusof, and Khalid}]{loghmanian2011nonlinear}
Loghmanian, S. M.~R., Yusof, R., and Khalid, M., \enquote{Nonlinear dynamic system identification using Volterra series: Multi-objective optimization approach,} \emph{2011 Fourth International Conference on Modeling, Simulation and Applied Optimization}, IEEE, 2011, pp. 1--5.

\bibitem[{Shiki et~al.(2023)Shiki, Hansen, and Silva}]{shiki2023practical}
Shiki, S.~B., Hansen, C., and Silva, S.~d., \enquote{Practical applications for nonlinear system identification using discrete-time Volterra series,} \emph{Journal of the Brazilian Society of Mechanical Sciences and Engineering}, Vol.~45, No.~2, 2023, p.~87.

\bibitem[{Lianrui and Ziniu(2023)}]{lianrui2023matched}
Lianrui, N., and Ziniu, W., \enquote{Matched Volterra reduced-order model for an airfoil undergoing periodic translation,} , 2023.

\bibitem[{Israelsen and Smith(2014)}]{israelsen2014generalized}
Israelsen, B.~W., and Smith, D.~A., \enquote{Generalized Laguerre reduction of the Volterra kernel for practical identification of nonlinear dynamic systems,} \emph{arXiv preprint arXiv:1410.0741}, 2014.

\bibitem[{De~Paula and Marques(2019)}]{de2019multi}
De~Paula, N., and Marques, F.~D., \enquote{Multi-variable Volterra kernels identification using time-delay neural networks: application to unsteady aerodynamic loading,} \emph{Nonlinear Dynamics}, Vol.~97, 2019, pp. 767--780.

\bibitem[{de~Paula et~al.(2019)de~Paula, Marques, and Silva}]{de2019volterra}
de~Paula, N.~C., Marques, F.~D., and Silva, W.~A., \enquote{Volterra kernels assessment via time-delay neural networks for nonlinear unsteady aerodynamic loading identification,} \emph{AIAA Journal}, Vol.~57, No.~4, 2019, pp. 1725--1735.

\bibitem[{Ross et~al.(2021)Ross, Smith, and {\'A}lvarez}]{ross2021learning}
Ross, M., Smith, M.~T., and {\'A}lvarez, M., \enquote{Learning Nonparametric Volterra Kernels with Gaussian Processes,} \emph{Advances in Neural Information Processing Systems}, Vol.~34, 2021, pp. 24099--24110.

\bibitem[{Wray and Green(1994)}]{wray1994calculation}
Wray, J., and Green, G.~G., \enquote{Calculation of the Volterra kernels of non-linear dynamic systems using an artificial neural network,} \emph{Biological cybernetics}, Vol.~71, No.~3, 1994, pp. 187--195.

\bibitem[{Levin et~al.(2022)Levin, Bastos, and Dowell}]{levin2022convolution}
Levin, D., Bastos, K.~K., and Dowell, E.~H., \enquote{Convolution and Volterra Series Approach to Reduced-Order Modeling of Unsteady Aerodynamic Loads,} \emph{AIAA Journal}, Vol.~60, No.~3, 2022, pp. 1663--1678.

\bibitem[{Balajewicz and Dowell(2012)}]{balajewicz2012reduced}
Balajewicz, M., and Dowell, E., \enquote{Reduced-order modeling of flutter and limit-cycle oscillations using the sparse Volterra series,} \emph{Journal of Aircraft}, Vol.~49, No.~6, 2012, pp. 1803--1812.

\bibitem[{Worden et~al.(1997)Worden, Manson, and Tomlinson}]{worden1997harmonic}
Worden, K., Manson, G., and Tomlinson, G., \enquote{A harmonic probing algorithm for the multi-input Volterra series,} \emph{Journal of Sound and Vibration}, Vol. 201, No.~1, 1997, pp. 67--84.

\bibitem[{Righi and Berci(2019)}]{righi2019subsonic}
Righi, M., and Berci, M., \enquote{On the Subsonic Lift of Elliptical Wings: Generation of Indicial Functions via CFD Simulations,} \emph{Journal of Aeroelasticity and Structural Dynamics}, Vol.~7, No.~1, 2019.

\bibitem[{Balajewicz et~al.(2010)Balajewicz, Nitzsche, and Feszty}]{balajewicz2010application}
Balajewicz, M., Nitzsche, F., and Feszty, D., \enquote{Application of multi-input Volterra theory to nonlinear multi-degree-of-freedom aerodynamic systems,} \emph{AIAA journal}, Vol.~48, No.~1, 2010, pp. 56--62.

\bibitem[{Cheng et~al.(2017)Cheng, Peng, Zhang, and Meng}]{cheng2017volterra}
Cheng, C., Peng, Z., Zhang, W., and Meng, G., \enquote{Volterra-series-based nonlinear system modeling and its engineering applications: A state-of-the-art review,} \emph{Mechanical Systems and Signal Processing}, Vol.~87, 2017, pp. 340--364.

\bibitem[{Campello et~al.(2004)Campello, Favier, and Do~Amaral}]{campello2004optimal}
Campello, R.~J., Favier, G., and Do~Amaral, W.~C., \enquote{Optimal expansions of discrete-time Volterra models using Laguerre functions,} \emph{Automatica}, Vol.~40, No.~5, 2004, pp. 815--822.

\bibitem[{Immordino et~al.(2023)Immordino, Da~Ronch, and Righi}]{immo2023ROM}
Immordino, G., Da~Ronch, A., and Righi, M., \enquote{Deep–learning framework for aircraft aerodynamics prediction,} \emph{AIAA AVIATION 2023 Forum}, 2023, p. 0179.

\bibitem[{Leishman(1993)}]{leishman1993indicial}
Leishman, J., \enquote{Indicial lift approximations for two-dimensional subsonic flow as obtained from oscillatory measurements,} \emph{Journal of Aircraft}, Vol.~30, No.~3, 1993, pp. 340--351.

\bibitem[{Economon et~al.(2016)Economon, Palacios, Copeland, Lukaczyk, and Alonso}]{economon2016su2}
Economon, T.~D., Palacios, F., Copeland, S.~R., Lukaczyk, T.~W., and Alonso, J.~J., \enquote{SU2: An open-source suite for multiphysics simulation and design,} \emph{Aiaa Journal}, Vol.~54, No.~3, 2016, pp. 828--846.

\bibitem[{Landon(1982)}]{landon1982naca}
Landon, R., \enquote{NACA 0012 oscillatory and transient pitching,} \emph{AGARD report}, Vol. 702, 1982, pp. 45--59.

\bibitem[{Yao and Liou(2016)}]{yao2016nonlinear}
Yao, W., and Liou, M.-S., \enquote{A nonlinear modeling approach using weighted piecewise series and its applications to predict unsteady flows,} \emph{Journal of Computational Physics}, Vol. 318, 2016, pp. 58--84.

\bibitem[{Piatak and Cleckner(2003)}]{piatak2003oscillating}
Piatak, D.~J., and Cleckner, C.~S., \enquote{Oscillating turntable for the measurement of unsteady aerodynamic phenomena,} \emph{Journal of aircraft}, Vol.~40, No.~1, 2003, pp. 181--188.

\bibitem[{Heeg et~al.(2016)Heeg, Chwalowski, Raveh, Jirasek, and Dalenbring}]{heeg2016overview}
Heeg, J., Chwalowski, P., Raveh, D.~E., Jirasek, A., and Dalenbring, M., \enquote{Overview and data comparisons from the 2nd Aeroelastic Prediction Workshop,} \emph{34th AIAA Applied Aerodynamics Conference}, 2016, p. 3121.

\bibitem[{Heeg and Piatak(2013)}]{heeg2013experimental}
Heeg, J., and Piatak, D.~J., \enquote{Experimental data from the benchmark supercritical wing wind tunnel test on an oscillating turntable,} \emph{54th AIAA/ASME/ASCE/AHS/ASC Structures, Structural Dynamics, and Materials Conference}, 2013, p. 1802.

\end{thebibliography}

\end{document}